\documentclass[aps,onecolumn]{revtex4}
\usepackage{epsfig}
\usepackage{graphicx}
\usepackage{amsmath,amssymb,amsfonts,bm}
\usepackage{array}
\usepackage{url}
\usepackage{hyperref}
\usepackage{multirow}
\usepackage{float}
\usepackage{lineno}
\usepackage{xspace}
\usepackage{ulem}
\usepackage[usenames,dvipsnames]{color}
\usepackage{epstopdf}
\begin{document}
	
	\title{Polarization Effects in Laser-Assisted (e,2e) Collision on H-atom by Twisted Electrons}
	\author{Neha}
	\email{p20210062@pilani.bits-pilani.ac.in}

	\author{Rakesh Choubisa}
	\email{rchoubisa@pilani.bits-pilani.ac.in }
	
	\affiliation{Department of Physics,  Birla Institute of Technology and Science, Pilani, Pilani Campus, Vidya Vihar, Pilani, Rajasthan 333031, India}

	\begin{abstract}
The dynamics of fast (e, 2e) collisions, induced by the impact of twisted electron beams, on atomic hydrogen, is analyzed in the presence of a laser field with circular and linear polarization. For the (e,2e) differential cross-section calculations we use Volkov and Coulomb-Volkov wave functions for scattered and ejected electrons, respectively, while the laser-atom interaction is treated in first-order perturbation theory. The formalism is developed for the asymmetric coplanar geometry in the first Born approximation. 
We investigate the influence of laser field polarization and provide a comparative analysis of Triple Differential Cross-Section's (TDCS) for circularly and linearly polarized laser fields as a function of ejected electron angle. The overall magnitude of the cross-section is
larger for circular polarization as compared to linear polarization. Some notable changes in the angular distributions of TDCS were also observed for the circular polarization as compared to that for linear polarization. 
 We further extend the study to coherent superpositions of twisted electron beams to examine OAM effects for macroscopic target which shows that the TDCS$_{av}$ is strongly sensitive to the difference of the projectile’s OAM as well as on the phase difference between them.

	\end{abstract}
	\date{\today}
	\maketitle

  \section{Introduction}
  \label{intro}
  The study of microscopic target ionization induced by electron impact, commonly known as the (e,2e) process, provides valuable insight into the electronic structure of the atomic target and the resulting residual ion \cite{byron1989theory}.
  Understanding of the electronic structure of microscopic targets is essential for modeling and analyzing laboratory plasmas, astrophysical processes, laser dynamics, and fields such as chemistry and biology \cite{bartschat2016electron,christophorou2000electron,shalenov2017scattering,de2019relativistic}. The field-free experimental studies have been conducted on various atomic targets \cite{ehrhardt1986differential,ehrhardt1969ionization,ehrhardt1982triply}.
  \\With the rapid development of laser technology and significant advances in multiparticle detection techniques, the study of laser-assisted electron–atom collision processes has attracted considerable experimental and theoretical interest. Electron-impact ionization of atoms in the presence of a laser field is commonly referred to as the laser-assisted (e,2e) collision \cite{ehlotzky1998electron}. 
  The laser-assisted electron-impact ionization experimental study was conducted on helium \cite{hohr2005electron,hohr2007laser}. 
  Several theoretical studies of triple differential cross-sections in different kinematical arrangements provide a general understanding of these phenomena \cite{mittleman2013introduction,francken1990theoretical,ehlotzky1998electron,ehlotzky2001atomic}. Initial studies of laser-assisted ionization processes did not account for the dressing effects of the target; however, these effects were included in the final state by representing them as Volkov or Coulomb-Volkov states \cite{MOHAN1978399,cavaliere1980particle, mandal1984electron,cavaliere1981effects,banerji1981electron,zangara1982influence,zarcone1983laser}. Joachain and his colleagues were the first to account for the dressing effects of the target due to the laser field \cite{joachain19882,martin1989electron}. They observed significant variations in the angular distributions of the differential cross-section. Numerous theoretical analyses on laser-assisted single and double ionization of atomic targets have been reported in the literature \cite{ghosh2009multiphoton,li2007ionization,li2005laser,chattopadhyay2005ionization,van2001double,sanz1999semiclassical,makhoute1999light,taieb1991light,khalil1997laser}.
  Taïeb et al. \cite{taieb1991light} and Makhoute et al. \cite{makhoute1999light} investigated the effects of laser polarization, specifically linear polarization ($\mathrm{LP}$) and circular polarization ($\mathrm{CP_Y}$), on the laser-assisted (e, 2e) process on hydrogen and helium, respectively. Their studies focused on angular distribution of TDCS as a function of ejected electron in an asymmetric coplanar geometry and demonstrated that a $\mathrm{CP_Y}$ laser produces a significantly larger TDCS compared to that for the $\mathrm{LP}$ lasers.
  These studies have opened new avenues for both theoretical and experimental research in this field. However, the reported studies were carried out using conventional electron beams described by plane waves, which do not carry orbital angular momentum. \\ 
 However, in the last few decades, the production of new structured light known as ``vortex beam or twisted beam" has led to a renewed interest in (e,2e) collisions for atomic and molecular targets \cite{harris2019ionization,harris2023controlling,dhankhar2020electron,dhankhar2022triple,dhankhar2022twisted,dhankhar2023dynamics,mandal2021semirelativistic,dhankhar2020double,van2014rutherford,van2015inelastic}.
     In particular, vortex beams have many properties that make them quite different from conventional plane waves. For example, a ``vortex electron beam" or a ``twisted electron beam" (TEB) carries quantized orbital angular momentum (OAM) along the propagation direction and has non-zero transverse linear momentum \cite{uchida2010generation}. The quantized OAM of TEB is a characteristic of the helical wavefronts possessed by the TEB \cite{lloyd2017electron,bliokh2017theory,larocque2018twisted}.
 The distinctive properties of twisted electron beams (TEBs),like orbital angular momentum and structured wavefront, provide new possibilities for fundamental studies and applications in areas such as high-resolution imaging, quantum state manipulation, optical trapping, information transfer, astronomy, and high-order harmonic generation \cite{verbeeck2010production,mcmorran2011electron,o2002intrinsic,larocque2018twisted,furhapter2005spiral,berkhout2009using,gemsheim2019high}. Collision physics, with its long-standing role in revealing fundamental aspects of atomic and molecular structure and few-body interactions, offers essential insights for understanding how structured electrons interact with matter. 
The above-discussed studies have been conducted for the field-free case. Our study aims to see the impact of the presence of a laser
field in asymmetric coplanar geometries.
In our previous work \cite{Neha2024},we investigated the laser assisted (e,2e) ionization of atomic hydrogen induced by twisted electron beams in the presence of a linearly polarized
laser field. In that study, the influence of the orbital angular momentum carried by the incident electron on the Triple Differential Cross Section (TDCS) was analyzed under well-defined kinematical conditions. 
Previous investigations of laser-assisted $(e,2e)$ processes with plane-wave projectiles \cite{taieb1991light,makhoute1999light} have reported angular distributions characterized by a dominant binary peak and a weaker, broadened recoil peak. These studies further showed that circularly polarized laser fields generally produce larger TDCS than that for linear polarization.
This enhancement has been associated with the distinct symmetry properties
and interaction dynamics introduced by circular polarization. 
The motivation of the present work is to extend our previous analysis to the case of circular polarization to examine whether the enhancement of the TDCS observed for circularly polarized fields persists when the incident electron is described by a twisted (vortex) wavefunction carrying orbital angular momentum. In addition to this we would like to compare the TDCS for the linearly and circularly polarised.
To provide a more experimentally relevant description, we also examine TDCS for the coherent superpositions of twisted electron beams, which enable a systematic investigation of the orbital angular momentum (OAM) effect on the TDCS. This allows us to explore how the phase structure of structured projectiles alters the collision dynamics during the laser-assisted ionization.
 
  The paper is organized as follows:
	In Section \ref{form}, we present our theoretical model, which is employed to study the laser-assisted (e, 2e) process for both plane wave and twisted electron beam impacts.
	In section \ref{result} we discusses our theoretical formalism.
	Section \ref{conc} presents our conclusions.\\
	Throughout the paper, atomic units are used unless stated otherwise.

	\section{\textbf{Theoretical Formalism}}
	\label{form}
    \subsection{Plane wave (e,2e) on H-atom}
The theoretical framework employed in the present work follows closely our
earlier formulation developed for the laser-assisted $(e,2e)$ ionization of
atomic hydrogen induced by twisted electron beams in a linearly polarized
laser field. In the present study, the formalism is extended to the case of a
circularly polarized laser field, while the treatment of the twisted
projectile and the target interaction remains unchanged.\\
The basic laser-assisted (e,2e) process accompanied by the absorption or emission of $l$ photons from the laser field can be described by the following equation:
       \begin{equation}\label{1}
		e^-(\mathbf{k_i}) + H + l\omega  \rightarrow  H^+ + e^-(\mathbf{k_s}) + e^-(\mathbf{k_e})
	\end{equation}
    Here,  $\boldsymbol{k_i}$, $\boldsymbol{k_s}$ and $\boldsymbol{k_e}$ incident momentum, scattered and ejected electron momentum respectively. $l$ is the number of photons exchanged between the laser-field and the atom-projectile system and $\omega$ is the photon energy. 
     In this study, we consider the laser field circularly polarised ($\mathrm{CP}$) and treat it classically  within the dipole approximation which means that it varies little over the atomic range. The vector potential of a $\mathrm{CP}$ field propagating along the $\hat{z}$-axis:
     \begin{equation}\label{2}
\mathbf{A}(t) \rightarrow A_{0}(\hat{x}  \ cos(\omega t) + \hat{y} \ sin(\omega t) \tan(\eta/2)) 
 \end{equation}
     the corresponding monochromatic electric field is:
      \begin{equation}\label{3}
\mathbf{E}(t) \rightarrow E_{0}(\hat{x}  \ sin(\omega t) - \hat{y} \ cos(\omega t) \tan(\eta/2)) 
 \end{equation}
     
      where $\mathbf{E_{0}}$ = $\mathbf{A_{0}}$ $\omega$/c  where, $E_{0}$ is the field strength, $\omega$ is the laser frequency, and $c$ is the speed of light. Here $\eta$ presents the degree of ellipticity.  Linear polarization corresponds to $\eta$ = 0 and circular polarization corresponds to $\eta$ = $\pi$/2. 
The main aim of this study is to calculate the transition matrix element ($T_{fi}$) in the first Born approximation (FBA) for the transition from the initial state $\psi_i$ to the final state $\psi_f$. For the laser-assisted (e,2e) process the transition matrix element in the first Born approximation is given as {\cite{joachain19882}}:
	\begin{equation}\label{4}
		T_{fi}^{B1}=-i\int_{-\infty}^{+\infty} dt\left< \chi _{k_{s}}\left ( \mathbf{r_{0}},t \right )\Phi_{k_{e}}\left ( \mathbf{r_{1}},t \right ) |\frac{-1}{r_0}  +  \frac{1}{|\mathbf{r_0} - \mathbf{r_i}|}| \chi _{k_{i}}\left ( \mathbf{r_{0}},t \right )\Phi_{0}\left ( \mathbf{r_{1}},t \right )\right>
	\end{equation}

     In equation \eqref{4}, $\mathbf{r_0}$ is the position vector of the incident (and scattered) electron, $\boldsymbol{r_i}$ represents the coordinates of the bound electron of the target.
   In equation \eqref{4}  $\chi _{k_{i}}$ and $\chi _{k_{s}}$ are the Volkov wavefunctions of the incident and scattered electron for the non-relativistic motion in circularly polarised laser-field and is described as  \cite{makhoute1999light} : 
   \begin{equation}\label{5}
     \chi_{\mathbf{k}}(\mathbf{r}_0,t)
= (2\pi)^{-3/2} \exp\!\left\{ i \left(\mathbf{k}\cdot \mathbf{r}_0- E_k t
- \Delta_{k}\sin(\omega t - \gamma_k) \right) \right\} 
   \end{equation}
 where, the effect of circularly polarised field is included by the term $\Delta_{k}$, given by equation \eqref{7}:
 \begin{equation}\label{6}
     \Delta_{k}
= \alpha_0({(\mathbf{k}\cdot \hat{\mathbf{x}})^2
+ (\mathbf{k}\cdot \hat{\mathbf{y}})^2 \tan^2\!\left(\frac{\eta}{2}\right)
})
 \end{equation}
 and
 \begin{equation}\label{7}
     \tan(\gamma_k)=\frac{\mathbf{k}\cdot \hat{\mathbf{y}}}
     {\mathbf{k}\cdot \hat{\mathbf{x}}}\,\tan\!\left(\frac{\eta}{2}\right)
 \end{equation}
The dressed state of the ejected electron in the continuum state under the influence of combined field of the residual ion and the external circularly polarised laser field, proposed by Joachain et al., \cite{joachain19882} can be written as :
\begin{equation}\label{8}
\begin{aligned}
\Phi_{k_e}^{(-)}(\mathbf{r}_1,t)
&= e^{-i \mathbf{a}\cdot \mathbf{r}_0}
   e^{-i E_{k_e} t}
   e^{-i \Delta_{k_e}\sin(\omega t - \gamma_{k_e})}
   \left\{
   \psi_{C,k_e}^{(-)}(\mathbf{r})
   \left[
   1 + i \Delta_{k_e}\sin(\omega t - \gamma_{k_e})
   \right] \right. \\
&\qquad \left.
   + \frac{i}{2}
   \sum_n
   \left[
   \frac{e^{i\omega t}}
        {E_n - E_{k_e} + \omega} - \frac{e^{-i\omega t}}
        {E_n - E_{k_e} - \omega}
   \right]
   M_{nj}\,
   \psi_n(\mathbf{r})
   \right\}
\end{aligned}
\end{equation}

    with
    \begin{equation}\label{9}
        M_{nj} = <\psi_n|\boldsymbol{\varepsilon}.\boldsymbol{r_i}|\psi_{k_e}>
    \end{equation}
  
Here $\psi_{C,k_{e}}^{-}( r_{1})$ is the Coulomb wave function, and $\boldsymbol{\varepsilon} = (\hat{\mathbf{e}}_{x} + i\,\hat{\mathbf{e}}_{y})/\sqrt{2}$ represents the polarization vector of the circularly polarized electric field, where $\hat{\mathbf{e}}_{x}$ and $\hat{\mathbf{e}}_{y}$ denote unit vectors along two mutually orthogonal directions in the polarization plane.

	Using wave functions of target, incident, scattered, and ejected electrons in the first Born $\mathit{T}$-matrix element (\eqref{4}) and integrating over time, we obtain $T_{fi}^{B1}$ for the circularly polarized light as:
	\begin{equation}\label{10}
		T_{fi}^{B1}=\left ( 2\pi  \right )^{-1} i\sum_{l=-\infty}^{+\infty}\delta \left (E_{k_{s}} + E_{k_{e}} - E_{k_{i}}- E_{0}-l\omega  \right )  e^{i l \gamma}f_{ion}^{B1,l}
	\end{equation}
    where 
    \begin{equation}\label{11}
\tan(\gamma)
=\frac{
-\Delta_{k_s}\sin(\gamma_{k_s})
+\Delta_{k_e}\sin(\gamma_{k_e})
-\Delta_{k_i}\sin(\gamma_{k_i})
}{
\Delta_{k_s}\cos(\gamma_{k_s})
+\Delta_{k_e}\cos(\gamma_{k_e})
+\Delta_{k_i}\cos(\gamma_{k_i})
}
\end{equation}

Within the first Born approximation, the scattering amplitude $f_{ion}^{B1,l}$ describes the laser-assisted ionization process accompanied by the net exchange of $l$ photons between the atomic system and the external circularly polarized laser field. Following the formalism of Ref.~\cite{joachain19882}, this amplitude naturally decomposes into three distinct contributions, reflecting the different interaction pathways induced by the presence of the circularly polarized electromagnetic field. In the following equations, we rewrite the decomposed form of the scattering amplitude for the circularly polarised field:
\begin{equation}\label{12}
f_{ion}^{B1,l}=f_I+f_{II}+f_{III}.
\end{equation}

	where, 
	\begin{equation}\label{13}
		f_{I}=-2\mathbf{\Delta^{-2}} J_{l}\left ( \lambda  \right ) \left<\psi_{C,k_{e}}^{-}\left|exp\left ( i\mathbf{\Delta} .\mathbf{r} - 1 \right ) \right|\Phi_{j} \right>,
	\end{equation}\\
	\begin{equation}\label{14}
		f_{II}=i\mathbf{\Delta^{-2}}\sum_{n}\left< \psi_{C,k_{e}}^{-}\left| \exp(i\mathbf{\Delta} .\mathbf{r} - 1)\right|\Phi_{n}\right>M_{nj}\left [ \frac{J_{l-1}\left ( \lambda  \right )}{E_{n}-E_{j}-\omega } -\frac{J_{l+1}\left ( \lambda  \right )}{E_{n}-E_{j}+\omega } \right ]
	\end{equation}\\
	\begin{equation}\label{15}
		\begin{aligned}
			f_{III}=i\mathbf{\Delta^{-2}}\sum_{n}\left< \Phi_{n}\left| \exp\left ( i\mathbf{\Delta} .\mathbf{r} - 1\right )\right|\Phi_{j}\right>M_{n,k_{e}}^{*}\left [ \frac{J_{l-1}(\lambda )}{E_{n}-E_{k_{e}}+\omega }-\frac{J_{l+1}\left ( \lambda  \right )}{E_{n}-E_{k_{e}}-\omega } \right ]\\
			-\frac{1}{2}\mathbf{\Delta_{k_{e}}} \left [J_{l-1} \left (\lambda   \right )  e^{i \gamma_{k_{e}}} -J_{l+1} \left ( \lambda  \right ) e^{-i \gamma_{k_e}}   \right ]\left< \psi_{C,k_{e}}^{-}\left| exp\left ( i\mathbf{\Delta} .\mathbf{r} \right )\right|\Phi_{j}\right>
		\end{aligned}
	\end{equation}
    Here, $J_{l}$ is the Bessel function of order $l$, $\lambda$, argument of the Bessel function, which represents the quiver motion of the charged particle in the external $\mathrm{CP}$ laser field, and is given as ; \cite{makhoute1999light}
    \begin{equation}\label{16}
\lambda^{\,2}
=\Delta_{k_s^{2}}+\Delta_{k_e^{2}}+\Delta_{k_i^{2}}
+2\Delta_{k_s k_e}\cos(\gamma_{k_s}-\gamma_{k_e})
-2\Delta_{k_s k_i}\cos(\gamma_{k_s}-\gamma_{k_i})
-2\Delta_{k_e k_i}\cos(\gamma_{k_e}-\gamma_{k_i}) .
\end{equation}

Within the framework of the first Born approximation, the triple differential cross section (TDCS) for the laser-assisted $(e,2e)$ process, accompanied by the absorption or emission of $l$ photons, is written as
\begin{equation}\label{17}
\frac{d^{3}\sigma^{\mathrm{B1},l}}{d\Omega_{s}\, d\Omega_{e}\, dE_{e}}
=
\frac{k_{s} k_{e}}{k_{i}}
\left| f_{\mathrm{ion}}^{\mathrm{B1},l} \right|^{2}
\end{equation}
where the scattering amplitude $f_{\mathrm{ion}}^{\mathrm{B1},l}$ is defined in Eq.~(12).

	\subsection{Twisted wave (e,2e) on H-atom}
	\label{TEB}
	
The theoretical framework employed in the present study is based on the approach developed in our previous study \cite{Neha2024} for the laser-assisted (e,2e) process induced by twisted electron beams (TEB) for the linearly polarised laser-field.
 A twisted electron beam (TEB) can generally be represented as a coherent superposition of tilted plane waves \cite{serbo2015scattering}. And for the laser-assisted processes, this description extends to a superposition of Volkov states \cite{karlovets2012electron}. The transition matrix element takes the following form; 
 \begin{equation}\label{18}
		T_{fi}^{tw}\left ( \varkappa,{\mathbf{\Delta}},\mathbf{b} \right )= \left ( -i \right )^{m_l}\sqrt{\frac{\varkappa
}{2\pi}}\int_{0}^{2\pi}\frac{d\phi_{p}}{2\pi}e^{im\phi_{p}-ik_{i\perp} \cdot {b}}T_{fi}\left ( \mathbf{\Delta} \right ),
	\end{equation}
    here momentum transfer $\mathbf{\Delta}$ depends on both longitudinal and transverse components of the incident beam \cite{harris2019ionization}. Here $T_{fi}(\Delta)$ is given by equation \eqref{10}. Here we compute the $T_{fi}(\Delta)$ for both the linearly and circularly polarised field. The transition amplitude in Eq.~\eqref{18} explicitly depends on the impact parameter $\boldsymbol{b}$, which specifies the position of the target relative to the axis of the incident twisted electron beam. In this communication, we extend this formalism for the circularly polarized laser field. This description corresponds to an idealized microscopic scenario in which the scattering occurs from a single, spatially localized atom or molecule with a well-defined impact parameter. However, such conditions are difficult to realize experimentally, and better ionization target would be macroscopic target which typically consist of a large ensemble of randomly distributed atoms or molecules within the interaction region. For a macroscopic target, the averaged cross section $TDCS_{av}$ is given by \cite{harris2019ionization}:

    \begin{equation}\label{19}
		TDCS_{av}=\frac{1}{2\pi\cos\theta_p}\int^{2\pi}_{0}d\phi_p \frac{d^3\sigma(\mathbf{\Delta})}{d\Omega_{e}d\Omega_{s}dE_{e}}.  
	\end{equation}
From \eqref{19}, we can see that the TDCS of a macroscopic target is independent on the projection of OAM ($m_l$) and the phase factor ($e^{im \phi}$). However, for such macroscopic targets we can recover the sensitivity of the TDCS to the OAM by treating the incident twisted electron beam as a coherent superposition of two twisted beams that share identical kinematic parameters but differ in their OAM projections.
The wave function of such a superposed twisted electron beam is given as	\cite{serbo2015scattering,karlovets2017scattering} :
\begin{equation}\label{20}
\Psi(\mathbf{r}) = c_1 \Psi_{m_{l_1}}(\mathbf{r}) + c_2 \Psi_{m_{l_2}}(\mathbf{r}) \,
\end{equation}
Here $\Psi_{m_{l}}(\mathbf{r})$ is wavefunction of the TEB (given in equation \eqref{20} in Ref \cite{Neha2024}), and the expansion coefficients $c_n$ given as;
\begin{equation}\label{21}
c_n = |c_n| e^{i \alpha_n}, \qquad |c_1|^2 + |c_2|^2 = 1 \, .
\end{equation}
When the incident projectile is described by the coherent superposition state given in Eq. \eqref{20}, the resulting TDCS depends on the interference of different OAMs and the relative phase between the two components. TDCS for the superposed twisted beam can be written in the following form;

\begin{equation}\label{22}
\mathrm{TDCS}_{\mathrm{av}}
= \frac{1}{2\pi \cos\theta_p}
\int_{0}^{2\pi} d\phi_p \,
G(\phi_p,\Delta m_l,\Delta \alpha)\,
\frac{d^3\sigma(\mathbf{\Delta})}{d\Omega_s \, d\Omega_e \, dE_e} \, .
\end{equation}
where $\Delta m_l = m_{l_2} - m_{l_1}$ is the difference in the OAM projections, 
$\Delta \alpha = \alpha_2 - \alpha_1$ is the difference in the phases of the twisted states, 
and the interference factor is given by 
$G(\phi_p,\Delta m,\Delta \alpha) = 1 + 2|c_1 c_2|\cos\!\left[(m_2 - m_1)\left(\phi_p - \frac{\pi}{2}\right) + \alpha_2 - \alpha_1\right]$.

  \section{Results and Discussions}
	\label{result}
   \begin{figure}
		\begin{tabular}{cc}
			\includegraphics[width=8.00cm]{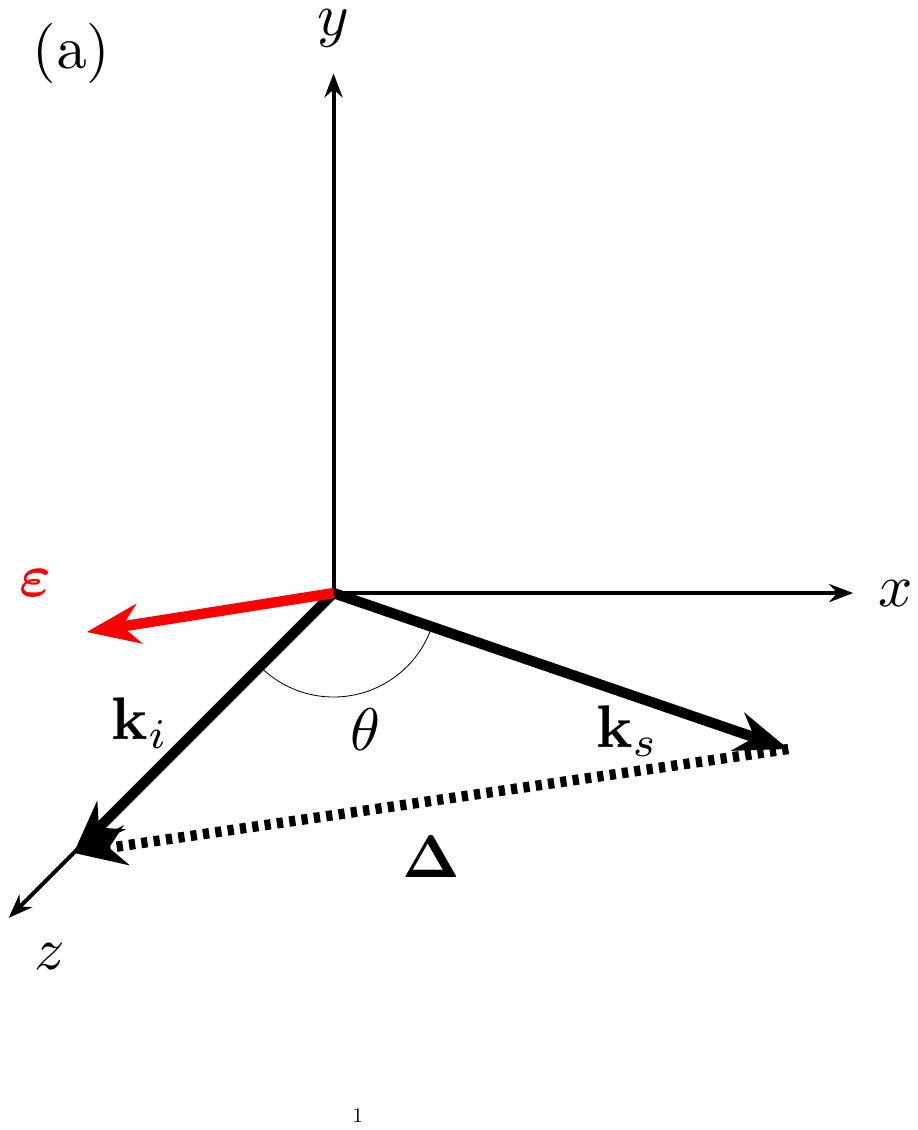}\ &
			\includegraphics[width=8.00cm]{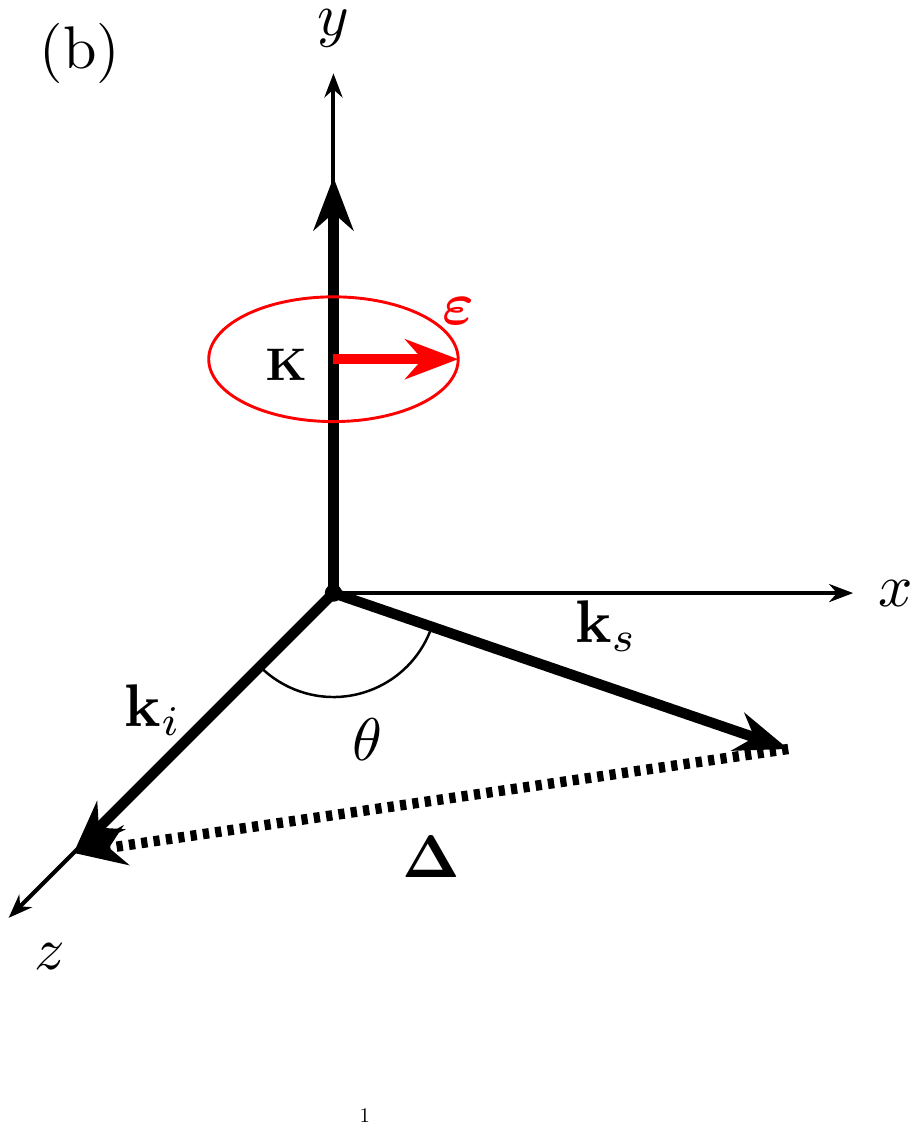}

		\end{tabular}
         \caption{Schematic illustration of the kinematic geometry for the laser-assisted (e,2e) reaction on atomic hydrogen. The incident electron momentum $\mathbf{k}_i$ is aligned along the $z$-axis, and $zx$  plane defines the scattering plane. (a) Configuration for linear polarization, where the electric field polarization vector $\boldsymbol{\varepsilon}$ is oriented parallel to the momentum transfer vector $\boldsymbol{\Delta}$. (b) Configuration for circular polarization ($\mathrm{CP_Y}$), where the laser field propagates along the $y$-axis ($\mathbf{k} \parallel \hat{y}$), such that the electric field vector $\boldsymbol{\varepsilon}$ rotates within the $zx$ scattering plane.}
		   \label{fig:geometry}
		\end{figure} 
	
This study investigates the ionization of atomic hydrogen for the laser-assisted (e,2e) process on H-atom utilizing both incident plane waves and Twisted Electron Beams (TEB). As depicted in Figure \ref{1}, the analysis concentrates on two distinct polarization states: (a) linear polarization, where the polarization vector $\boldsymbol{\varepsilon}$ aligns parallel to the momentum transfer vector $\boldsymbol{\Delta}$ and (b) circular polarization ($\mathrm{CP_Y}$), characterized by the propagation along the Y-axis with the field polarized within the $zx$ scattering plane. 
 

	\subsection{Angular profile of TDCS for the laser-assisted (e,2e) process} 
\begin{figure}
    \centering
    \includegraphics[width=0.5\linewidth]{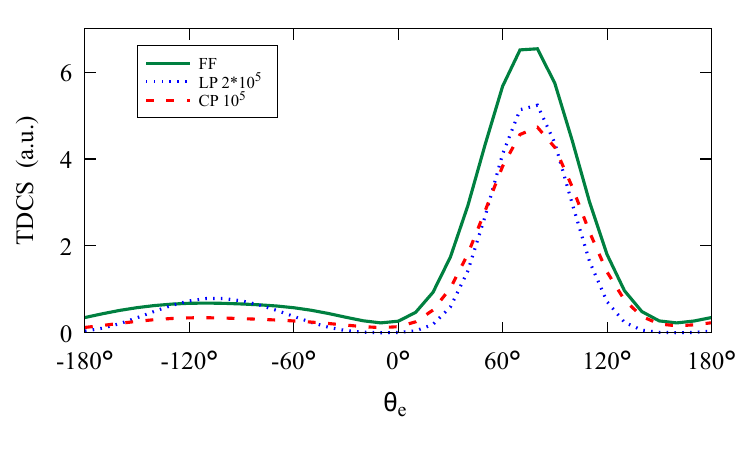}
    \caption{Angular distributions of the Triple Differential Cross Section (TDCS) plotted against the ejected electron angle $\theta_e$. The study compares three interaction regimes: the Field Free ($\mathrm{FF}$) case (solid curve), laser-assisted collisions in a Linearly Polarized ($\mathrm{LP}$) field (blue dotted curve), with the polarization vector parallel to the momentum transfer direction; and laser-assisted collisions in a Circularly Polarized ($\mathrm{CP_Y}$) field (red dashed curve), with the polarization vector aligned along the $y$-axis within the scattering ($zx$) plane. Kinematic parameters are fixed at incident energy $E_i = 600$ eV, ejected energy $E_e = 5$ eV, and scattering angle $\theta_s = 4^\circ$. Laser parameters correspond to the absorption of $l = 1$ photon with field strength $\varepsilon = 10^6$ V/m and photon energy $\hbar\omega = 1.17$ eV. The specific scaling factors are indicated in the panel to compare with the field free results.}
    \label{fig:2}
\end{figure}
Figure~\ref{fig:2} shows the angular distribution of the triple differential cross section (TDCS) for laser-assisted ionization of atomic hydrogen in the Ehrhardt asymmetric coplanar geometry, calculated within the first Born approximation for an incident electron energy $E_i = 600$ eV, scattering angle $\theta_s = 4^\circ$, and fixed ejected electron energy $E_e = 5$ eV, as a function of the ejection angle $\theta_e$. The laser field corresponds to single-photon absorption ($l = 1$) with field strength $\varepsilon = 10^6$ V/m and photon energy $\hbar\omega = 1.17$ eV. Results obtained for circular polarization ($\mathrm{CP_Y}$) are compared with those for linear polarization with the polarization vector parallel to the momentum transfer direction ($\varepsilon \parallel \Delta$). The calculations reproduce the characteristic binary and recoil peak structures reported previously \cite{taieb1991light}, with a dominant binary peak and a reduced, broadened recoil peak. Moreover, the TDCS magnitude for circular polarization is found to be approximately twice that obtained for the corresponding linear polarization case, in agreement with earlier theoretical predictions \cite{makhoute1999light}. The overall agreement with the available data confirms the validity of the present approach in the high-impact-energy regime, where exchange effects are negligible and laser-induced modifications dominate over higher-order contributions.
			
		\begin{figure*}[htp!]
				
				\begin{tabular}{ccc}
					\includegraphics[width=6.00cm]{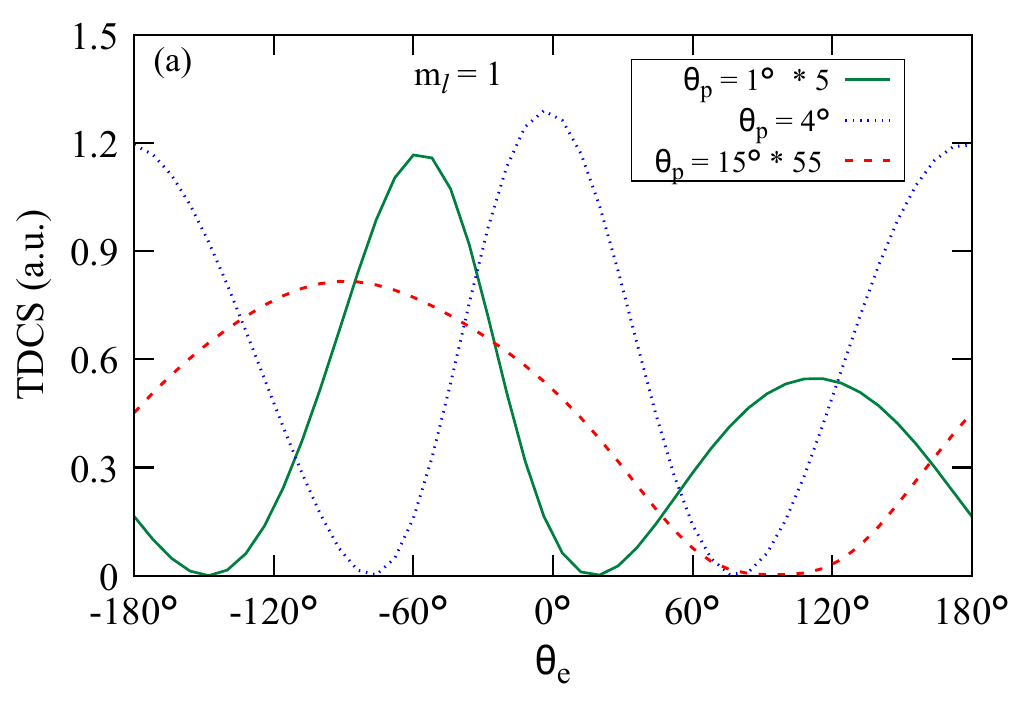}\ & \includegraphics[width=6.00cm]{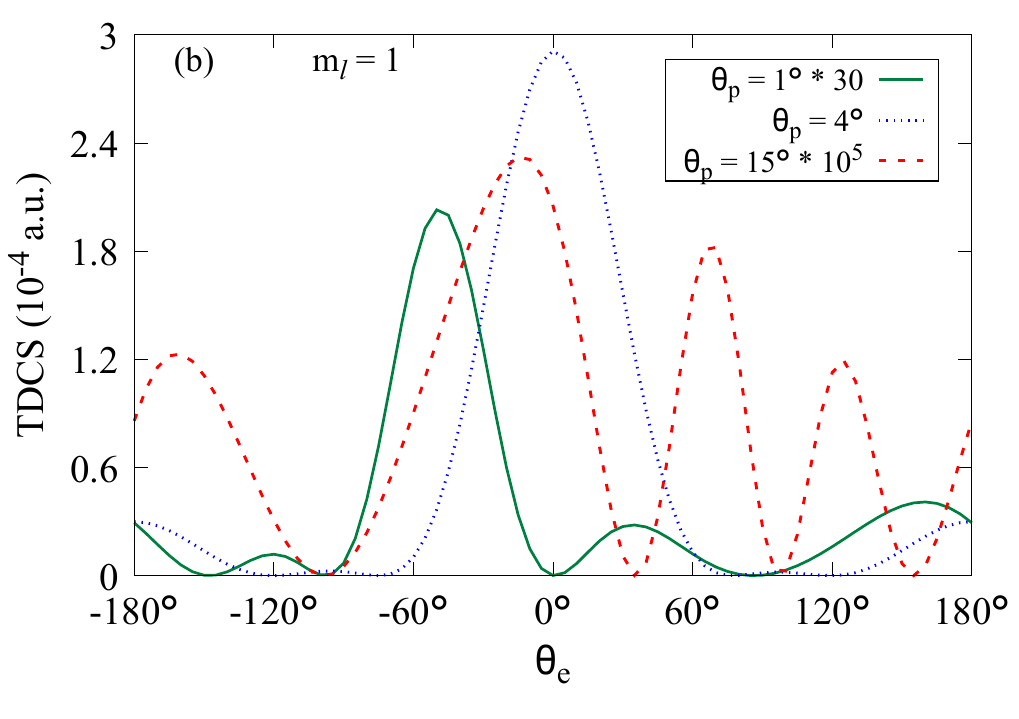}\ &
					\includegraphics[width=6.00cm]{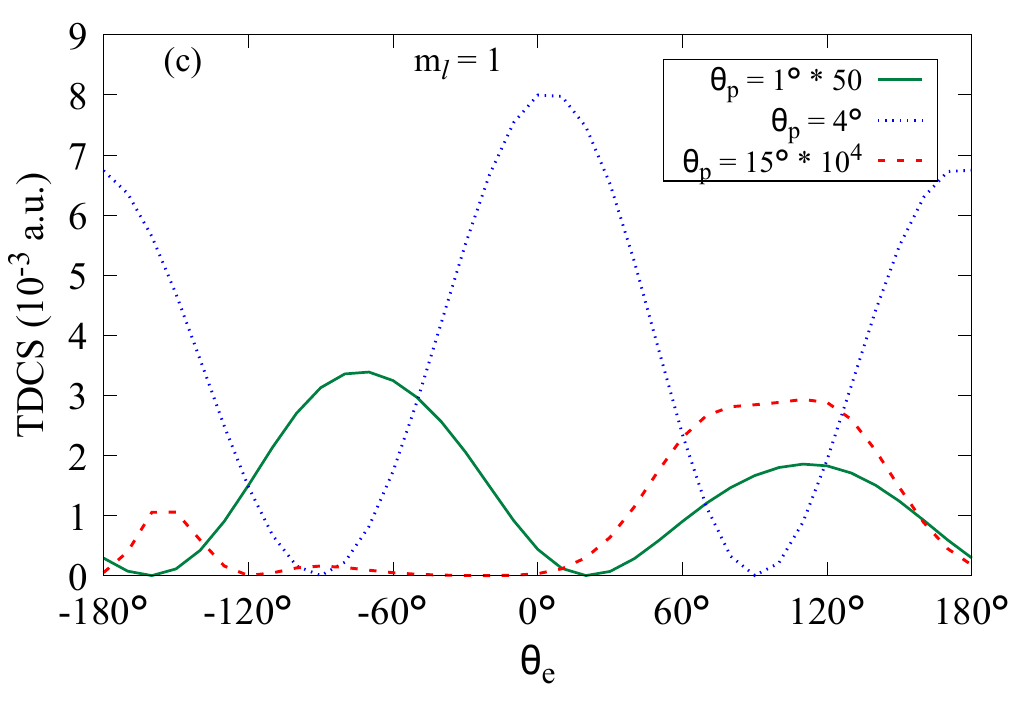}
					
				\end{tabular}

				\caption{Angular distributions of the Triple Differential Cross Section (TDCS) for orbital angular momentum $m_l$ = 1, plotted against the ejected electron angle $\theta_e$. The study compares three interaction regimes: (a) the Field Free ($\mathrm{FF}$) case; (b) laser-assisted collisions in a Linearly Polarized ($\mathrm{LP}$) field, with the polarization vector parallel to the momentum transfer direction; and (c) laser-assisted collisions in a Circularly Polarized ($\mathrm{CP_Y}$) field, with the polarization vector aligned along the $y$-axis within the scattering ($zx$) plane. The curves denote opening angles of $\theta_p = 1^\circ$ (solid green), $\theta_p = 4^\circ$ (blue dotted), and $\theta_p = 15^\circ$ (red dashed). Kinematic parameters are same as the fig \ref{1}.}   
			\label{fig:3}
			\end{figure*}
		
In figure \ref{fig:3} we present the TDCS for TEB collision as a function of ejected electron angle $\theta_e$ for the kinematics as figure \ref{2} keeping OAM $m_l$ = 1 and selecting different $\theta_p$ as mentioned in the graph. In figures \ref{fig:3}(a), (b), and (c), we illustrate the variations of the TDCS for opening angle $\theta_p$ = 1$^\circ$ (solid green curve), 4$^\circ$ (blue dotted curve), and 15$^\circ$ (red dashed curve). Here we compare the results for field free case ($\mathrm{FF}$), linearly polarised field ($\mathrm{LP}$) and circularly polarised laser field ($\mathrm{CP_Y}$). In a quantitative comparison of the TDCS magnitudes for all three scenarios ($\mathrm{FF}$, $\mathrm{LP}$, and $\mathrm{CP_Y}$), we observed that the $\mathrm{FF}$ process yields the maximum amplitude. When a laser field is introduced,  a significant suppression is observed: the $\mathrm{CP_Y}$ field attenuates the TDCS magnitude by approximately two orders of magnitude relative to the $\mathrm{FF}$ study. This effect is even more pronounced in the linearly polarized ($\mathrm{LP}$) case, where the cross-section diminishes by roughly three orders of magnitude (see figure \ref{fig:3}(a),(b) and (c)). These findings demonstrate that the ionization probability is also highly sensitive to the specific polarization state, similar to the plane wave case in the figure \ref{2}. \\
 A comparative analysis of the angular distribution of TDCS reveals that the observed behavior is highly sensitive to the relation between the scattering angle $\theta_s$
 and the opening angle $\theta_p$ of the twisted electron beam. We observe from all the plots that TDCS exhibit the maximum magnitude for
 $\theta_s$ =  $\theta_p$, relative to the kinematics when  $\theta_s$ $\neq$  $\theta_p$.\\
In particular, for the specific kinematic condition  $\theta_s$ =  $\theta_p$, the angular distribution obtained in the $\mathrm{CP_Y}$ case seems to follow roughly similar patterns that of the $\mathrm{FF}$ process. Here we observed two peak structures, a forward peak in the direction $\theta_e$ = 0$^\circ$ and backward peak in the $\theta_e$ = $\pm$180$^\circ$ direction  (see blue dotted curve in figure \ref{fig:3}(a) and (c)).   
In contrast, for the $\mathrm{LP}$  field, the backward peak in the $\theta_e$ = $\pm$180$^\circ$ is suppressed, and a prominent forward peak in the direction $\theta_e$ = 0$^\circ$ is observed (see blue dotted curve in figure \ref{fig:3}(b)). 
The similarity between the $\mathrm{CP_Y}$ and $\mathrm{FF}$ angular distributions is not a general feature. When $\theta_s$ $\neq
$ $\theta_p$, the angular distribution for the $\mathrm{CP_Y}$ and the $\mathrm{FF}$ case deviates significantly from one another (see the green solid and red dashed curves in figure \ref{fig:3}(a) and (c)).
This similarity between the $\mathrm{FF}$ and $\mathrm{CP_Y}$ angular behavior observed for $\theta_s$ =  $\theta_p$ =  4$^\circ$ (in this study)
may be understood in terms of an effective kinematic symmetry associated with the momentum structure of the twisted electron beam. A twisted beam consists of a coherent superposition of plane-wave components whose momenta lie on a cone defined by the opening angle $\theta_p$.
 When the scattering angle is equal to the opening angle, contributions from the dominant transverse momentum components of the incident beam symmetrically enter the scattering process. Since a circularly polarized laser field does not introduce a fixed direction in the scattering plane, this effective symmetry is preserved, leading to an angular distribution that closely resembles the field-free case.

 \begin{figure*}[htp!]
				
				\begin{tabular}{ccc}
					\includegraphics[width=6.00cm]{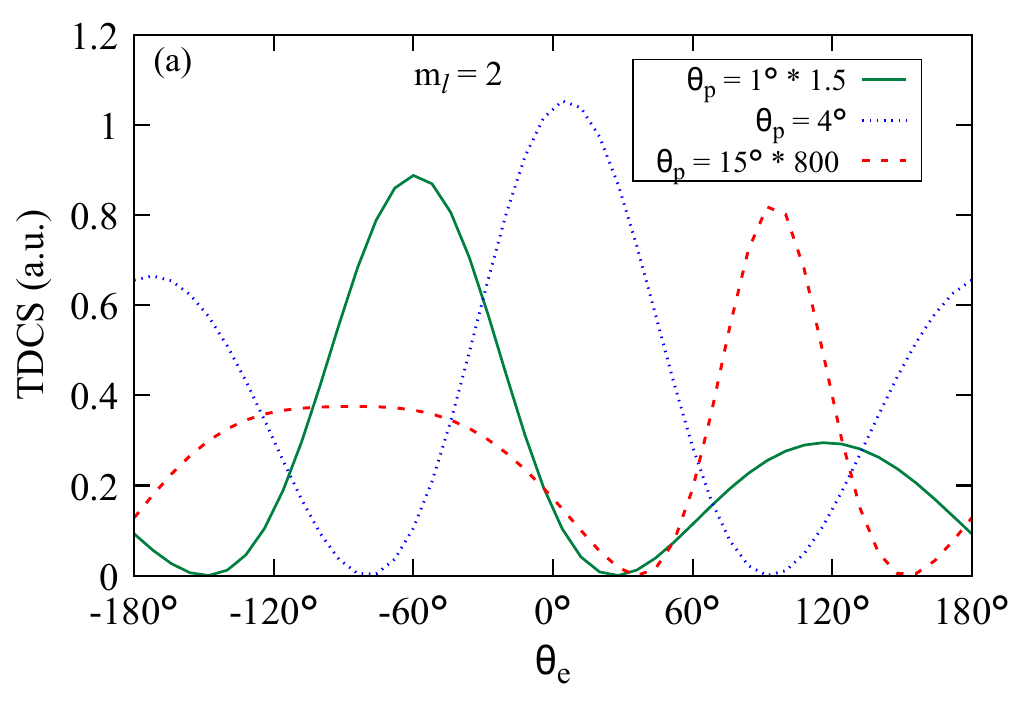}\ & \includegraphics[width=6.00cm]{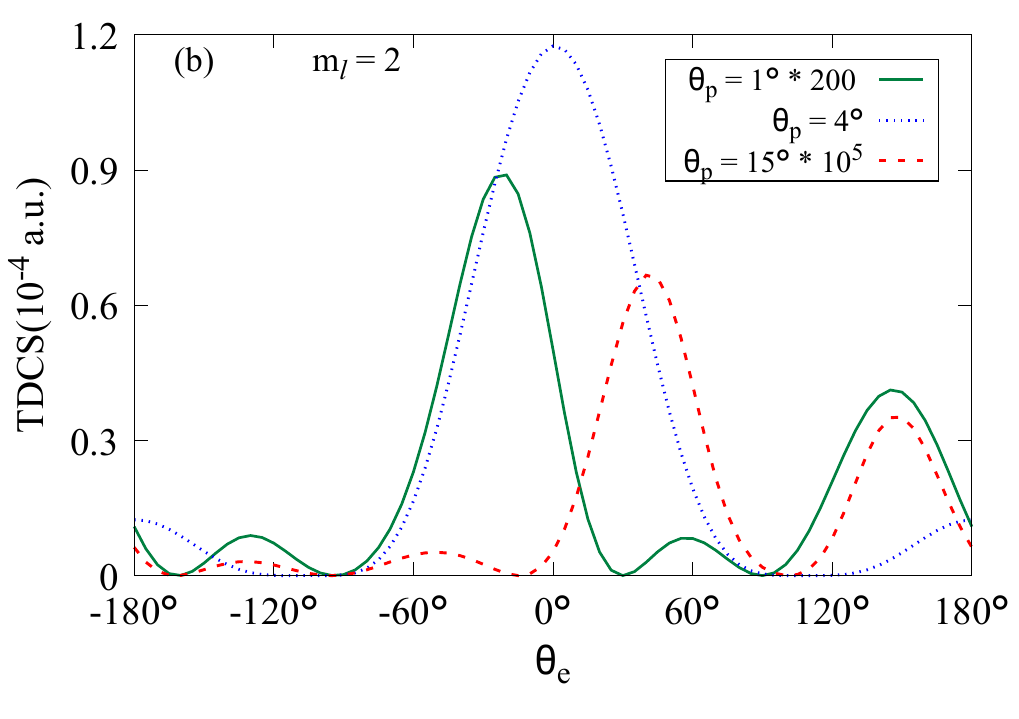}\ &
					\includegraphics[width=6.00cm]{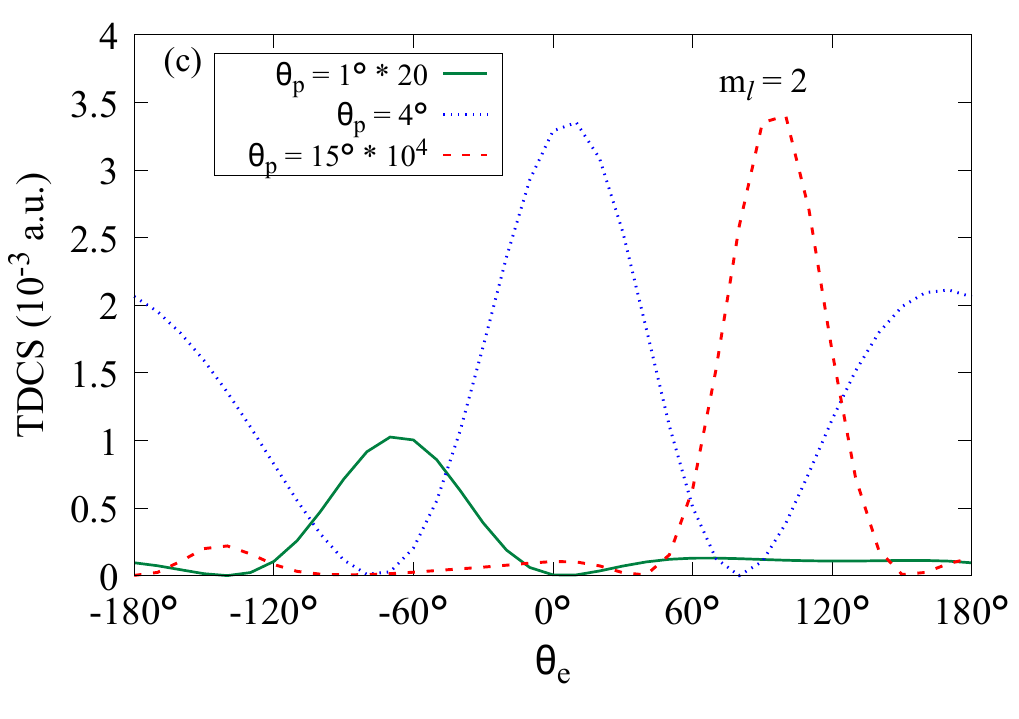}
					
				\end{tabular}

				\caption{Kinematics is same as figure \ref{fig:3} except the $m_l$ = 2 }   
			\label{fig:4}
			\end{figure*}
    Figure \ref{fig:4} presents the TDCS for $m_l$ = 2, with the same kinematics employed in figure \ref{fig:3}, except OAM ($m_l$ = 2). Here we observe that with the increment of OAM the magnitude of the TDCS decreases.\\ In particular, compared to the field-free case, the magnitude of TDCS is suppressed by two orders of magnitude in the presence of a $\mathrm{CP_Y}$ field and by about three orders of magnitude for the $\mathrm{LP}$ field ($\mathrm{LP}$).
For the specific kinematic condition $\theta_s = \theta_p = 4^\circ$, the
angular distributions for the $\mathrm{FF}$ and $\mathrm{CP_Y}$ cases again
exhibit similar features, characterized by the presence of both forward and
backward emission peaks. This behavior contrasts with that observed for the
$\mathrm{LP}$ case, where the backward emission is strongly suppressed and the TDCS is dominated by a forward peak. This trend is consistent with the behavior observed for the $m_l$ = 1 and confirms that the similarity between the $\mathrm{FF}$ and $\mathrm{CP_Y}$ cases persists at $m_l = 2$ only for this specific geometrical condition.

When the scattering angle differs from the opening angle, namely for
$\theta_p = 1^\circ$ and $15^\circ$, the angular distributions for the three scenarios ($\mathrm{FF}$, $\mathrm{LP}$, and $\mathrm{CP_Y}$) differ significantly, as shown in Figs.~\ref{fig:4}(a)--(c) (see green solid and red dashed curves). In the field-free case [Fig.~\ref{fig:4}(a)], the TDCS exhibits two distinct peaks at $\theta_e = -60^\circ$ and $120^\circ$ for $\theta_p = 1^\circ$ (see green solid curve), whereas for $\theta_p = 15^\circ$ the distribution is characterized by a pronounced peak near $\theta_e = 90^\circ$ together with a broad structure extending from $\theta_e = 0^\circ$ to $180^\circ$ (see red dashed curve).

In contrast, the laser-assisted cases display a simpler angular behavior. For
both $\mathrm{LP}$ and $\mathrm{CP_Y}$, the TDCS is dominated by a single main peak when $\theta_s \neq \theta_p$. In the $\mathrm{LP}$ case
[Fig.~\ref{fig:4}(b)], the TDCS exhibits a prominent peak located near
$\theta_e = -30^\circ$ for $\theta_p = 1^\circ$ (see green solid curve) and near
$\theta_e = 40^\circ$ for $\theta_p = 15^\circ$, accompanied by weaker
subsidiary lobes (see red dashed curve).
For the circularly polarized case $\mathrm{CP_Y}$ ( Fig.~\ref{fig:4}(c)), a similar single-peak
structure is observed, but without the additional small lobes present in the
$\mathrm{LP}$ distributions. At $\theta_p = 1^\circ$, the TDCS is characterized by a dominant peak near $\theta_e = -60^\circ$ (see green solid curve), while for
$\theta_p = 15^\circ$ the main peak shifts to $\theta_e = 90^\circ$ (see red dashed curve). The
absence of secondary lobes in the $\mathrm{CP_Y}$ case further emphasizes the
role of laser polarization in determining the detailed angular structure of the TDCS.

         \begin{figure*}[htp!]
				
				\begin{tabular}{ccc}
					\includegraphics[width=6.00cm]{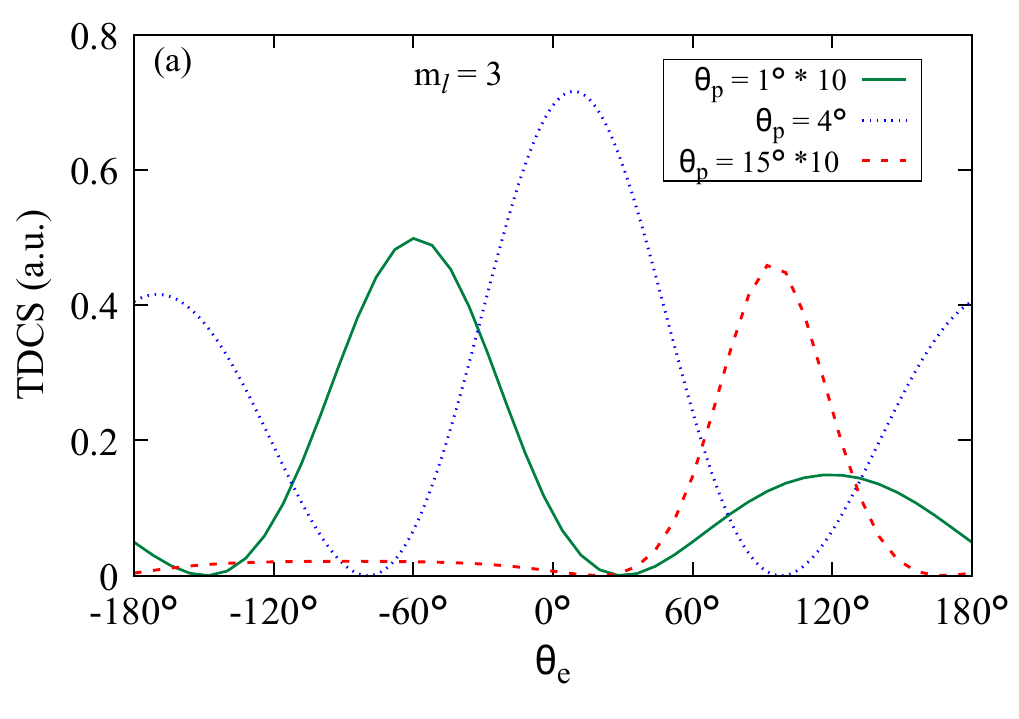}\ & \includegraphics[width=6.00cm]{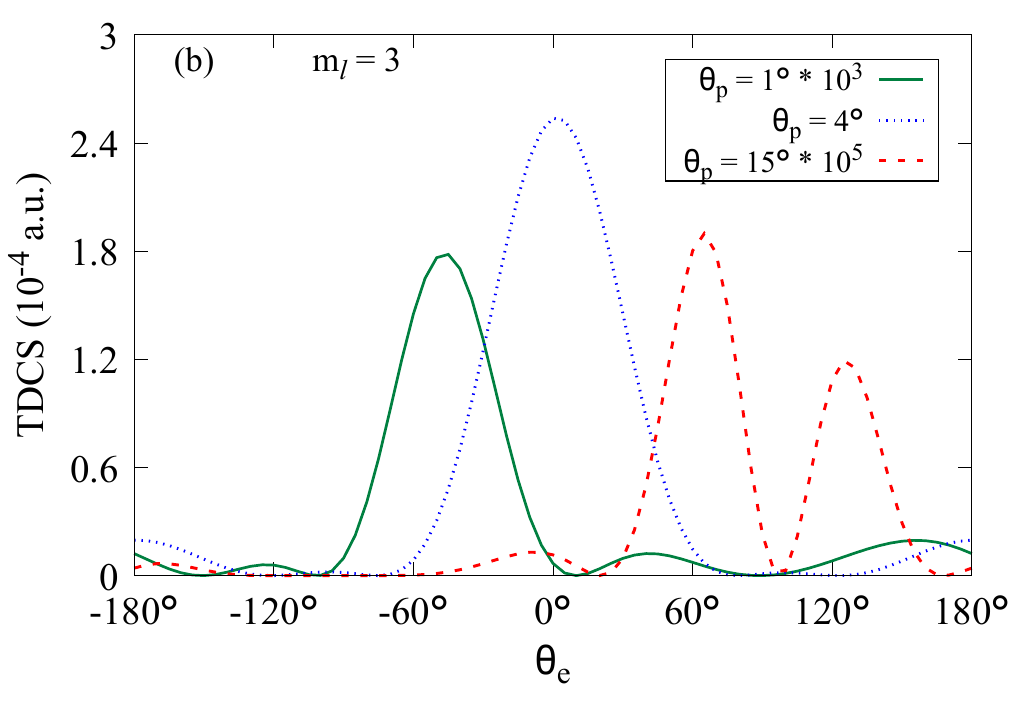}\ &
					\includegraphics[width=6.00cm]{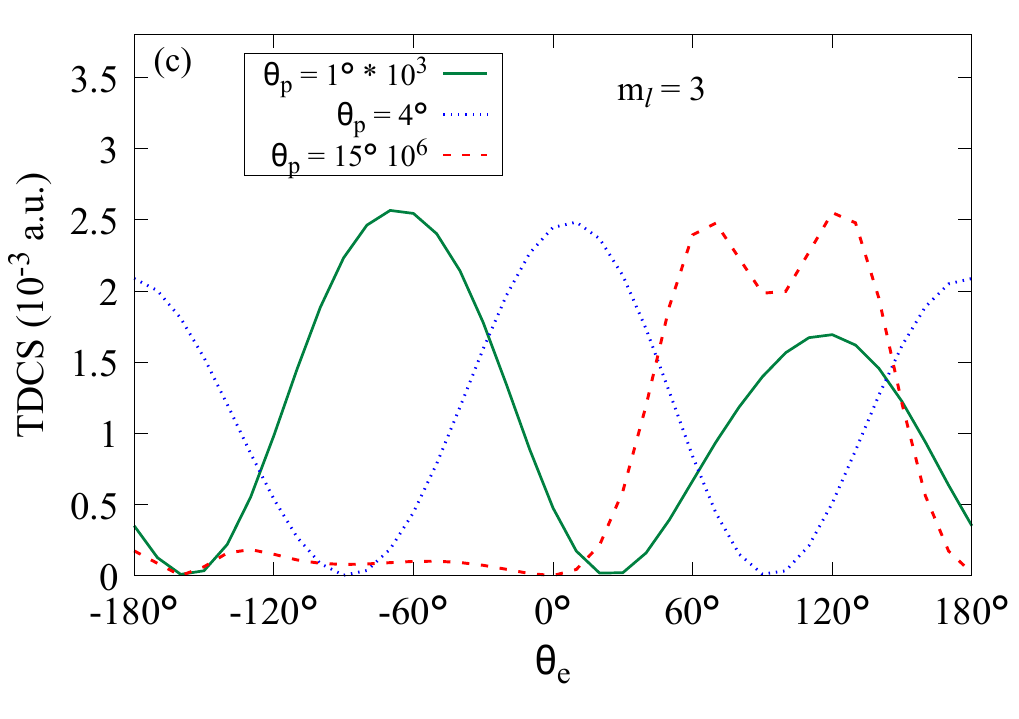}
					
				\end{tabular}

				\caption{Kinematics is same as figure \ref{fig:3} except the $m_l$ = 3 }   
			\label{fig:5}
			\end{figure*}
\begin{figure*}[htp!]
				
				\begin{tabular}{ccc}
					\includegraphics[width=6.00cm]{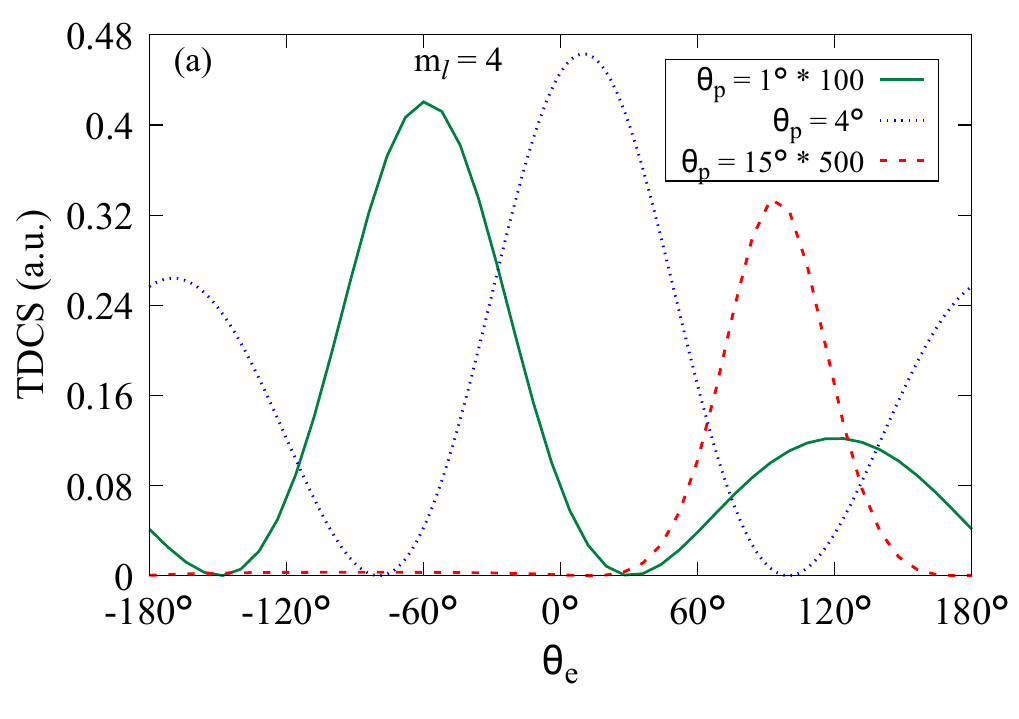}\ & \includegraphics[width=6.00cm]{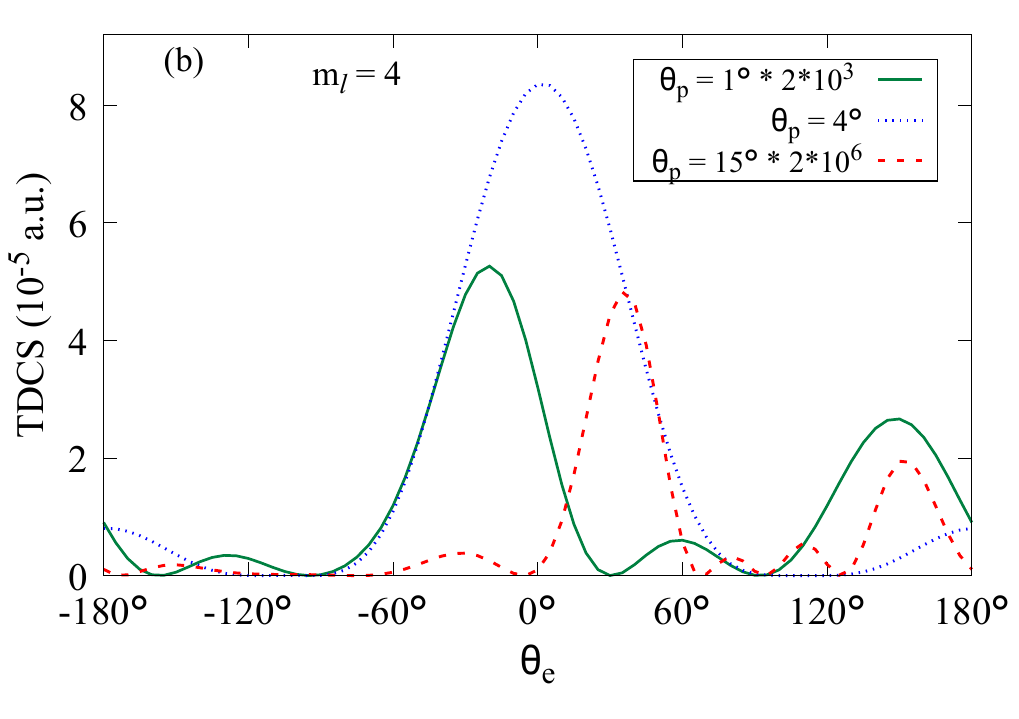}\ &
					\includegraphics[width=6.00cm]{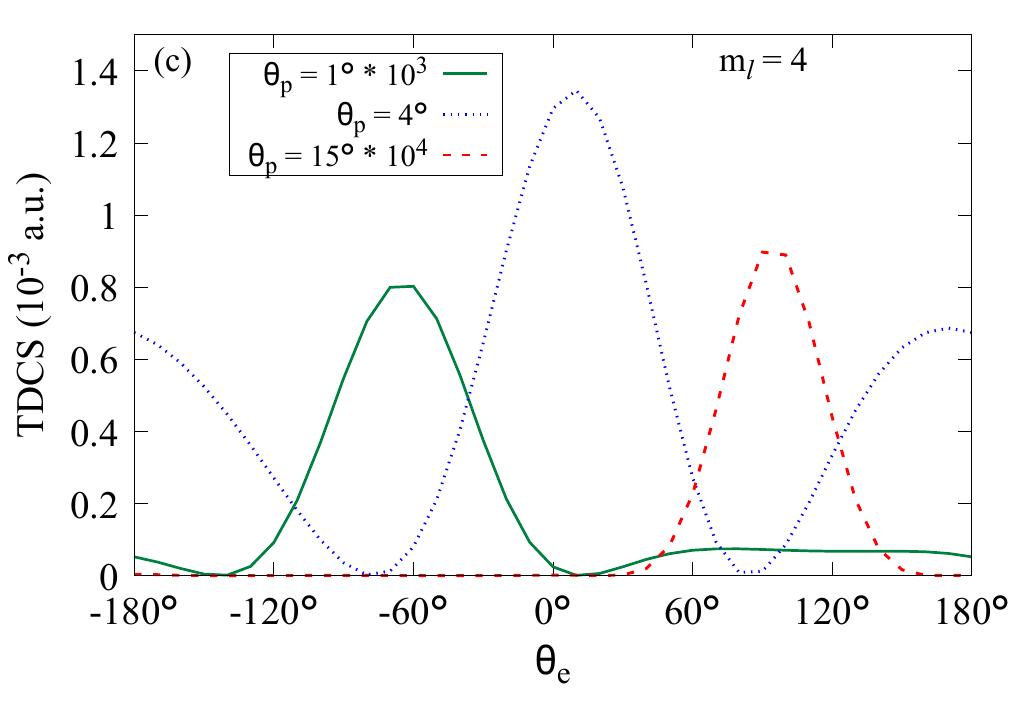}
					
				\end{tabular}

				\caption{Same as figure \ref{fig:3} except $m_l = 4$}   
			\label{fig:6}
			\end{figure*}

  
          An analysis of the TDCS for higher OAM's  $m_l = 3$ and $4$, presented in Figs.~\ref{fig:5} and \ref{fig:6} reveal a further reduction in magnitude compared to those obtained for $m_l = 1$ and $2$ shown in Figs.~\ref{fig:3} and \ref{fig:4}. This progressive decrease in the TDCS with increasing OAM is consistent with the general trend observed throughout the present study, indicating that higher OAM states of the twisted projectile lead
to a reduced ionization probability.
Regarding the angular distribution, we observe that for the kinematic condition $\theta_s = \theta_p$, the TDCS profiles for $m_l = 3$ and $4$ exhibit qualitative features similar to those of $m_l = 1$ and $2$. This angular distribution similarity persists across all three interaction regimes investigated: Field-Free ($\mathrm{FF}$), Linear Polarization ($\mathrm{LP}$), and Circular Polarization ($\mathrm{CP_Y}$). This observation confirms that the geometric symmetry between the scattering angle and the opening angle (i.e. $\theta_s = \theta_p$) plays a dominant
role in shaping the angular structure of the TDCS.\\
In the $\mathrm{FF}$ scenario, a detailed comparison shows that the angular distribution remains similar at small projectile scattering angles ($\theta_p = 1^\circ$ and $4^\circ$) for all OAM ($m_l$) values considered in this work. However, significant deviations arise when the opening
angle is increased to $\theta_p = 15^\circ$,  indicating an enhanced sensitivity to the transverse momentum content of the twisted beam at larger opening angles.
Upon the introduction of the laser field, the scattering dynamics exhibit a distinct dependence on the OAM's odd-even nature.
For the kinematic case $\theta_s = \theta_p$, the angular distribution remains uniform across all studied OAM values for  $\mathrm{LP}$ and  $\mathrm{CP_Y}$  polarization conditions. In contrast, when $\theta_p = 1^\circ$, a clear dependence on the nature of the OAM becomes apparent in the laser-assisted cases. Odd OAM values ($m_l = 1$ and $3$) yield nearly identical angular distributions, characterized by a dominant peak around $\theta_e = 50^\circ$. In contrast, even OAM values ($m_l = 2$ and $4$) exhibit a peak shifted toward $\theta_e = 30^\circ$, accompanied by small lobes, as illustrated by the solid green curves in Figs.~\ref{fig:3}--\ref{fig:6}. This odd–even character of the OAM-dependent behavior highlights the role of the phase structure of the twisted projectile in the presence of the laser field. \
Similar to the $\mathrm{FF}$ case, this odd–even symmetry breaks down at $\theta_p = 15^\circ$, where the angular distribution varies differently with the OAM number (red dashed curves in Figs. \ref{fig:3}-\ref{fig:6}). The angular distributions thus show a clear dependence on the nature of $m_l$, where odd values exhibit similar profiles that differ from those corresponding to even values. These results suggest that the combined effects of the laser polarization, the orbital angular momentum of the twisted electron beam, and the beam opening angle influence the angular structure of the TDCS.

           \subsection{Triple Differential Cross-Sections for macroscopic Target}
\begin{figure*}[htp!]
				
				\begin{tabular}{ccc}
					\includegraphics[width=6.00cm]{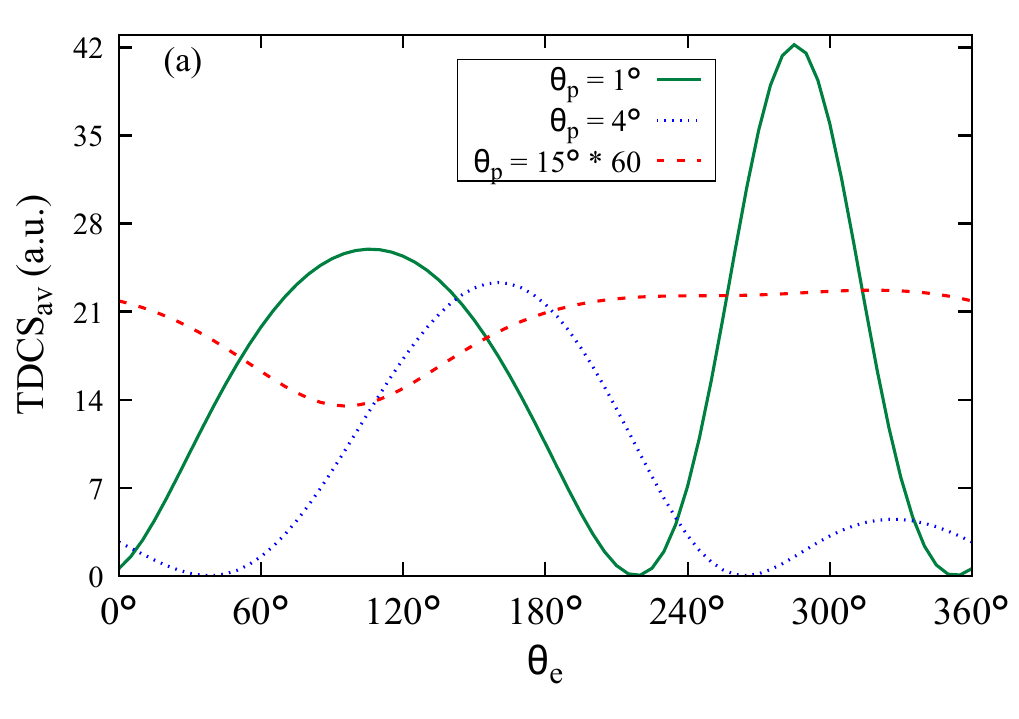}\ & \includegraphics[width=6.00cm]{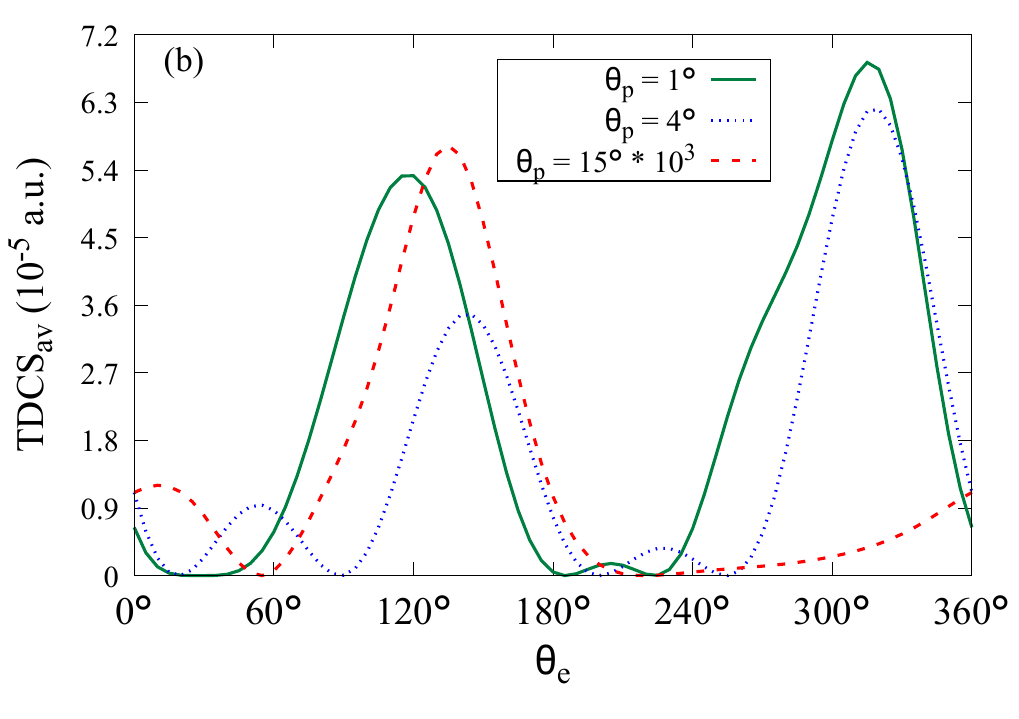}\ &
					\includegraphics[width=6.00cm]{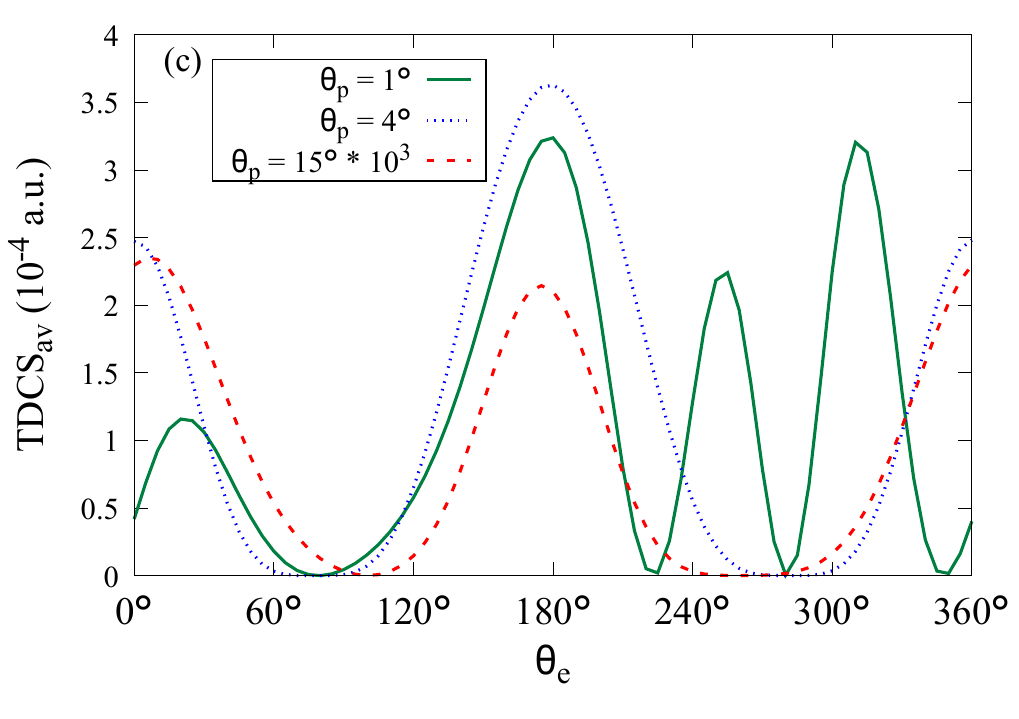}
					
				\end{tabular}

				\caption{Triple Differential Cross Section of macroscopic target ($TDCS_{av}$) for hydrogen atoms by TEB, plotted against the ejected electron angle $\theta_e$. The study compares three interaction regimes: (a) the Field Free ($\mathrm{FF}$) case; (b) laser-assisted collisions in a Linearly Polarized ($\mathrm{LP}$) field, with the polarization vector parallel to the momentum transfer direction; and (c) laser-assisted collisions in a Circularly Polarized ($\mathrm{CP_Y}$) field, with the polarization vector aligned along the $y$-axis within the scattering ($zx$) plane. The curves denote opening angles of $\theta_p = 1^\circ$ (solid green), $\theta_p = 4^\circ$ (blue dotted), and $\theta_p = 15^\circ$ (red dashed).The kinematics employed is same as of figure \ref{fig:3}. The specific scaling factors are indicated in the respective panels.}  
			\label{fig:7}
			\end{figure*}

  Here we present the angular distribution $TDCS_{av}$ as a function of $\theta_e$ for the macroscopic target, which represents the realistic experimental scenario of twisted electrons interacting with a macroscopic target. Figure \ref{fig:7} illustrate $TDCS_{av}$ for three scenarios (a) for field free $\mathrm{FF}$, (b) for laser-assisted with linearly polarised laser-field $\mathrm{LP}$ and (c) for circularly polarised laser field $\mathrm{CP_Y}$. $TDCS_{av}$ is calculated at the same kinematics employed in the Figure \ref{fig:3}. Here we like to retreat that the $TDCS_{av}$ only depends on opening angle and independent on OAM. The opening angles considered here are as follows $\theta_p$ = 1$^\circ$ , 4$^\circ$ and 15$^\circ$, and the corresponding TDCS$_{av}$ angular profiles are represented by green solid curve, blue dotted curve and red dashed curve respectively.
  A detailed comparison of the, $TDCS_{av}$, for the three cases considered in this work—namely the field-free (FF) case [Fig.~\ref{fig:7}(a)], linear polarization (LP) [Fig.~\ref{fig:7}(b)], and circular polarization ($\mathrm{CP_Y}$) [Fig.~\ref{fig:7}(c)]—reveals that the presence of the laser field significantly modifies both the magnitude and the angular distribution of $TDCS_{av}$.  
 For the field-free case at an opening angle $\theta_p = 1^\circ$, two pronounced peaks are observed in the vicinity of the perpendicular emission directions,
$\theta_e = 90^\circ$ and $270^\circ$. This symmetry is altered when the
laser field is introduced. In the $\mathrm{LP}$ case, the angular distribution
exhibits two shifted maxima located at $\theta_e = 120^\circ$ and $310^\circ$. In contrast, for the $\mathrm{CP_Y}$ configuration, the distribution is dominated by a single peak centered around $\theta_e = 180^\circ$, accompanied by additional weaker side peaks.  
These observations clearly demonstrate that the polarization state of the laser field
plays a crucial role in reshaping the ionization dynamics and the angular emission
pattern of electrons from macroscopic targets.
From our results we also observed that the peak value of the $TDCS_{av}$, decreases as the opening angle of the twisted electron beam increases to $\theta_p$ = 15$^\circ$ (a significant reduction in the magnitude can be seen from figure \ref{fig:7} as red dashed curve is scaled up by different factor for three different case for better comparison). This trend is consistent with our earlier findings obtained from the TDCS calculations in figures \ref{fig:3}-\ref{fig:6}.
            This reduction in the magnitude can be understood as follows; the opening angle $\theta_p$ determines how the projectile's total momentum is divided into transverse ($k_\perp = k \sin\theta_p$) and longitudinal ($k_z = k \cos\theta_p$) components. Increasing $\theta_p$ naturally increases the transverse momentum while reducing the forward (longitudinal) momentum. Consequently, the effective electron flux available for ionization in the scattering direction plane. Furthermore, a larger opening angle expands the cone of plane-wave components in momentum space.
            This combination of reduced longitudinal momentum leads to a systematic reduction in the magnitude of $TDCS_{av}$ and TDCS as the opening angle increases.

 \subsection{Triple differential cross-section for superposition of two Bessel Beams}

  \begin{figure*}[htp!]
				
				\begin{tabular}{ccc}
					\includegraphics[width=6.00cm]{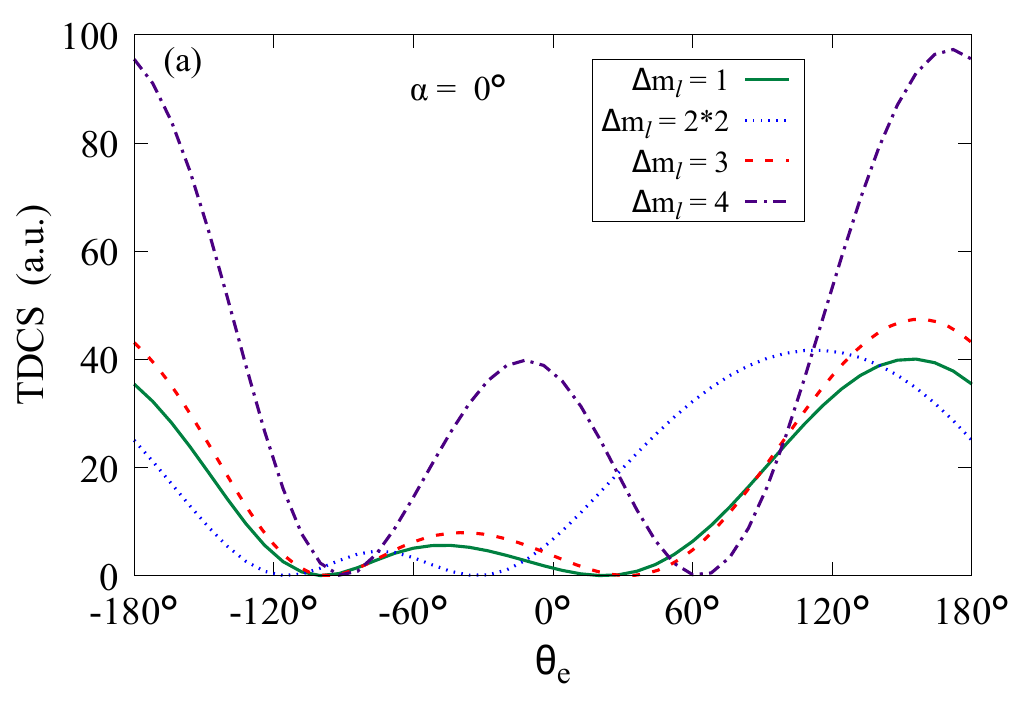}\ & \includegraphics[width=6.00cm]{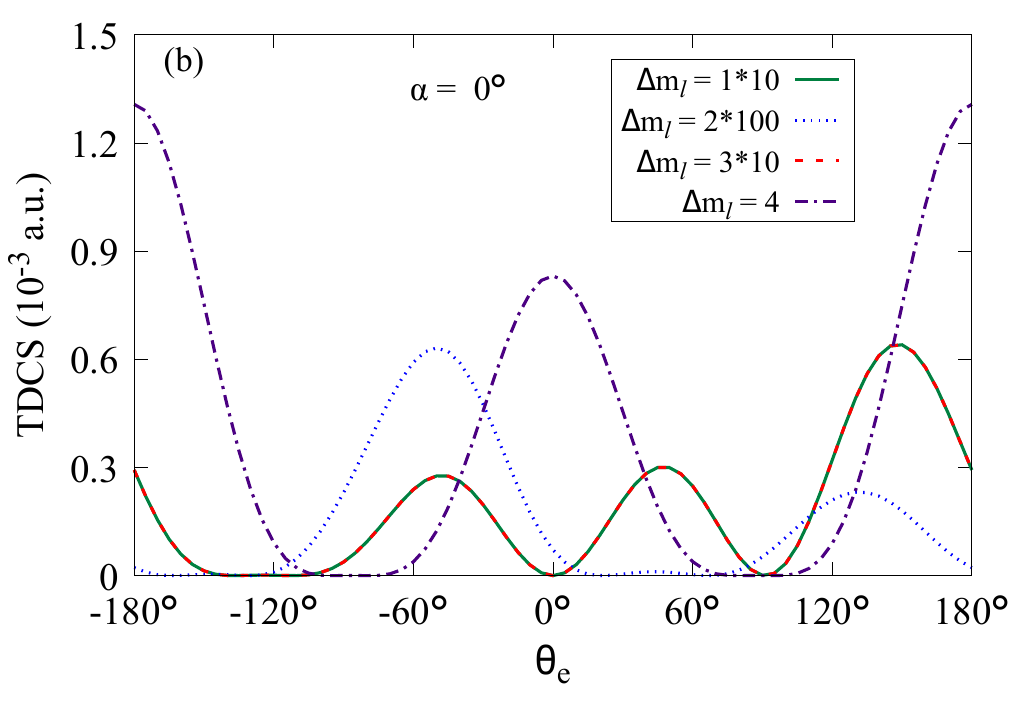}\ &
					\includegraphics[width=6.00cm]{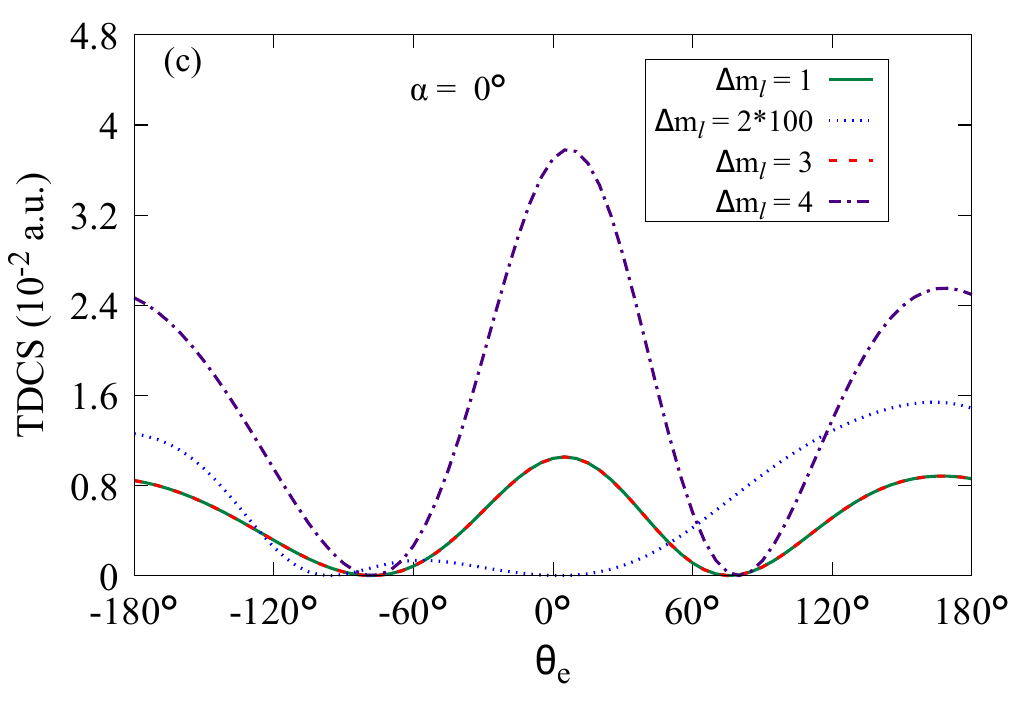}
					
				\end{tabular}

				\caption{Triple Differential Cross Section (TDCS) for hydrogen atoms by coherent superposition of two Bessel beams for phase difference $\alpha$ = 0$^\circ$.
                 The study compares three interaction regimes: (a) the Field Free ($\mathrm{FF}$) case; (b) laser-assisted collisions in a Linearly Polarized ($\mathrm{LP}$) field,and (c) laser-assisted collisions in a Circularly Polarized ($\mathrm{CP_Y}$) field. Kinematics employed is same as of figure \ref{fig:3}. The different OAM projections for which calculations are performed are mention in the penal.}    
			\label{fig:8}
			\end{figure*}
\begin{figure*}[htp!]
				\begin{tabular}{ccc}
					\includegraphics[width=6.00cm]{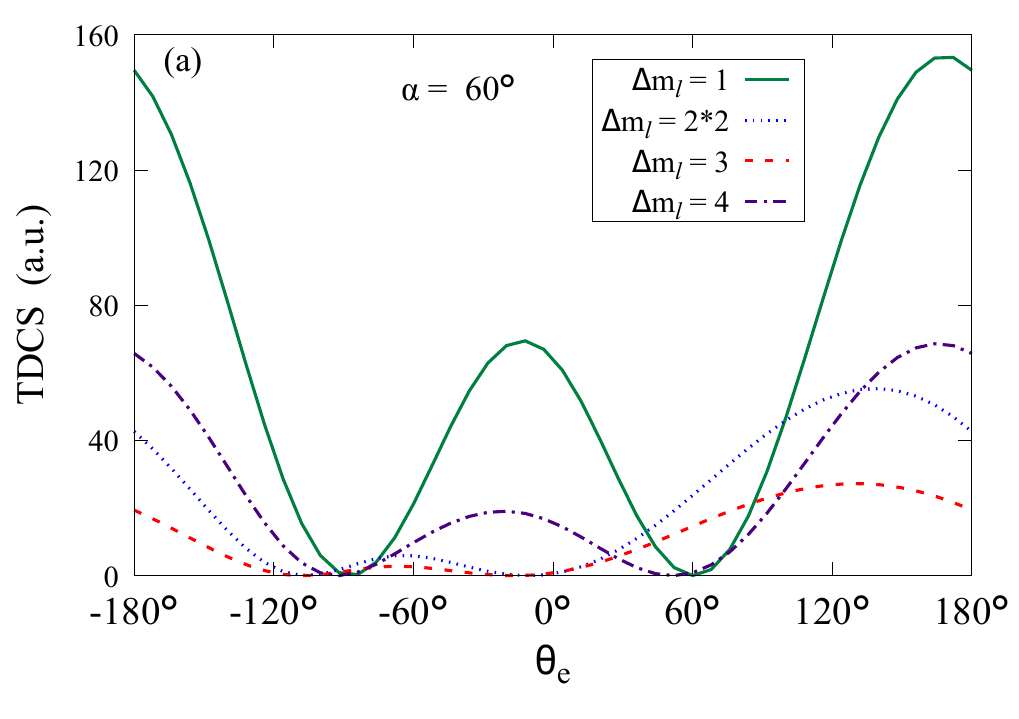}\ & \includegraphics[width=6.00cm]{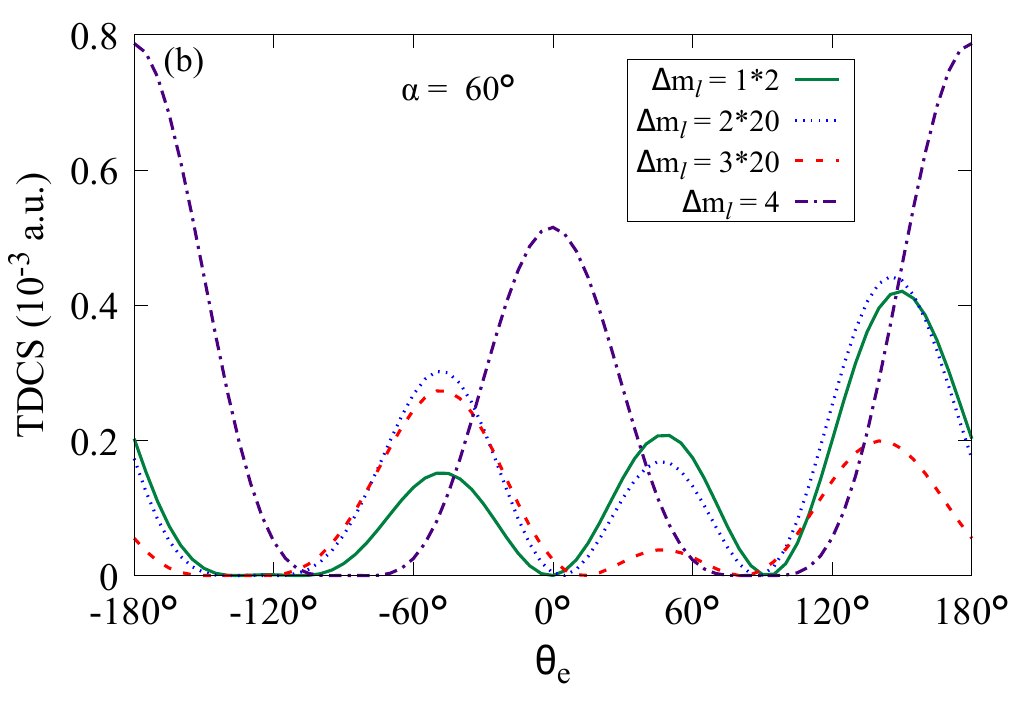}\ &
					\includegraphics[width=6.00cm]{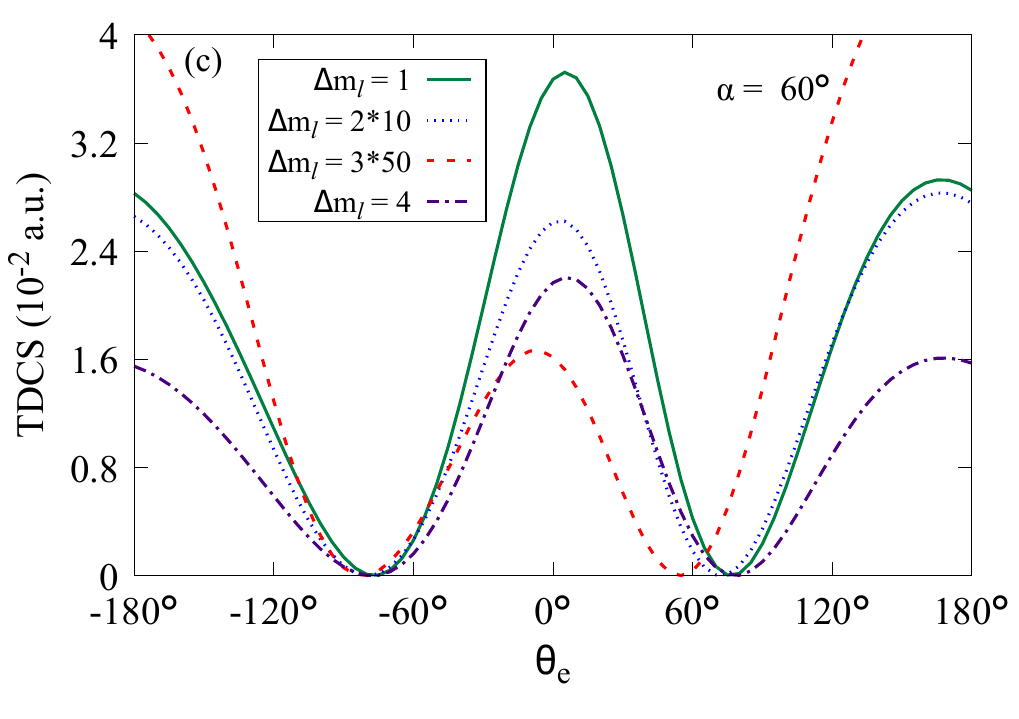}
					
				\end{tabular}

				\caption{Same as figure \ref{fig:8} except phase factor $\alpha$ = 60$^\circ$}   
			\label{fig:9}
			\end{figure*}

\begin{figure*}[htp!]
				
				\begin{tabular}{ccc}
					\includegraphics[width=6.00cm]{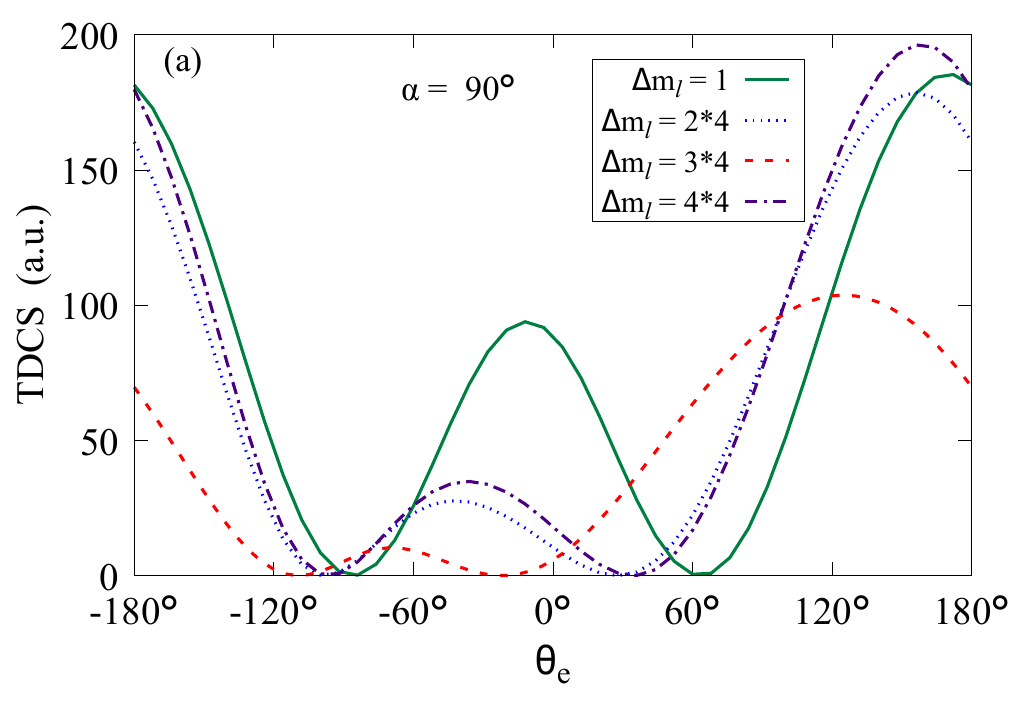}\ & \includegraphics[width=6.00cm]{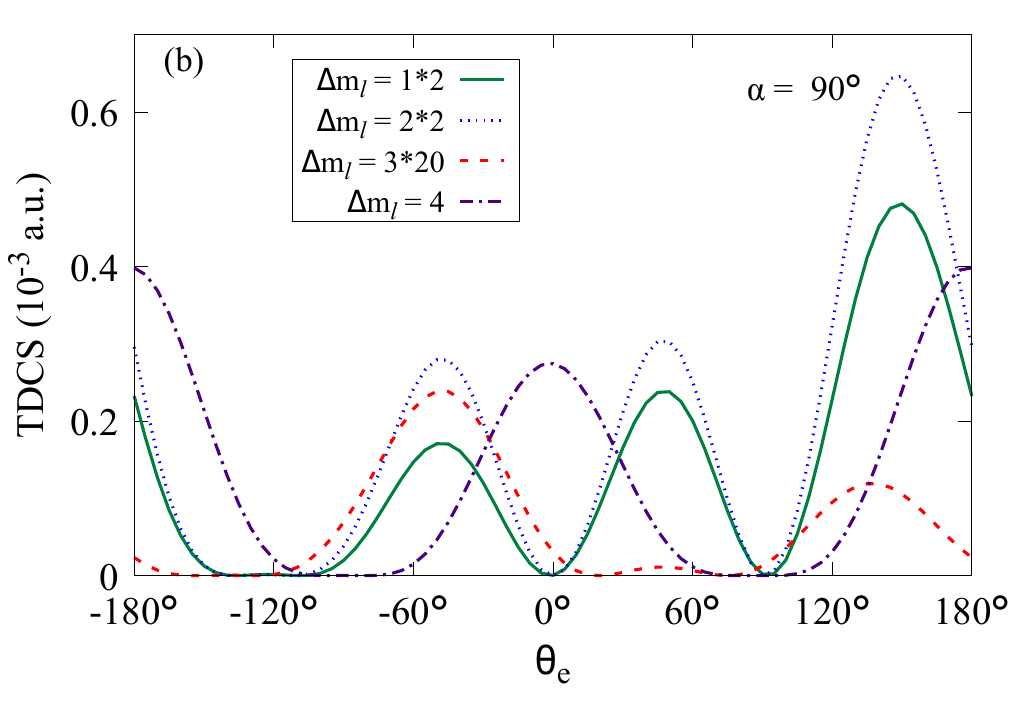}\ &
					\includegraphics[width=6.00cm]{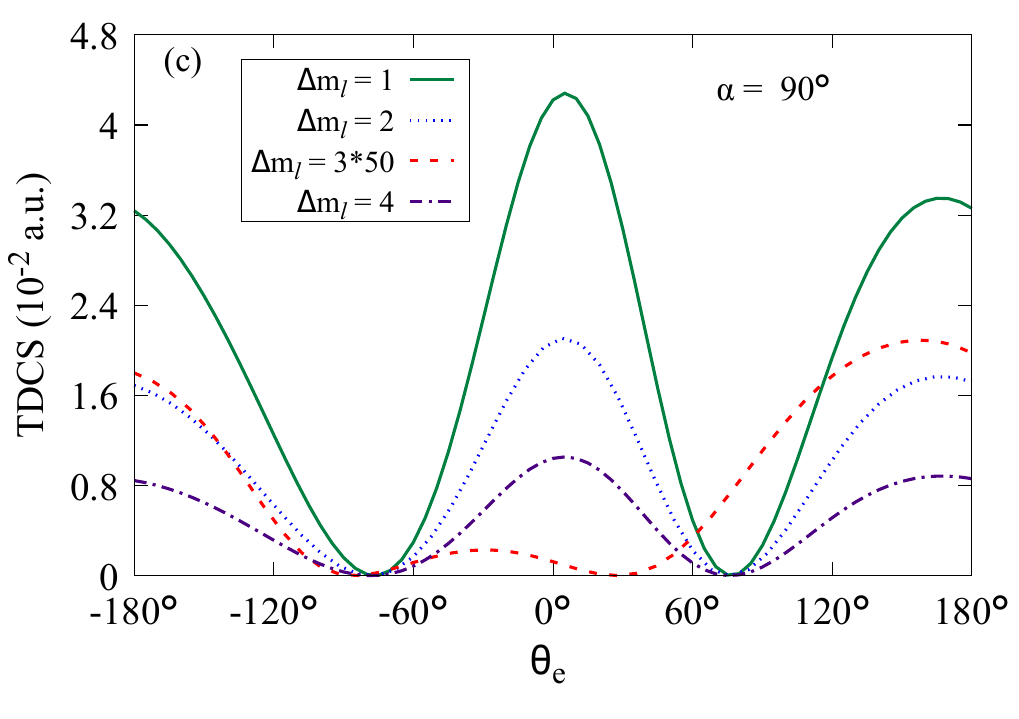}
					
				\end{tabular}

				\caption{Same as \ref{fig:8} except phase factor $\alpha$ = 90$^\circ$. }   
			\label{fig:10}
			\end{figure*}

 
         Figures~\ref{fig:8}–\ref{fig:10} present the TDCS for the coherent superposition of two
Bessel beams for relative phases $\alpha = 0^\circ$, $60^\circ$, and $90^\circ$. The use of a superimposed Bessel-beam configuration provides a realistic framework for examining
the influence of the OAM projection $\Delta m_l$ on the $(e,2e)$ process. The TDCS is
evaluated for $\Delta m_l = 1, 2, 3,$ and $4$, and are represented by the green solid, blue dotted, red dashed, and purple dash–dotted curves, respectively. The results are shown for
(a) the field-free case (FF), (b) a linearly polarized laser field (LP), and
(c) a circularly polarized laser field ($\mathrm{CP_Y}$), under the same kinematical
conditions as those employed in Figure~\ref{fig:7}.

For $\alpha = 0^\circ$, corresponding to the case where the two incident twisted electron
beams are in phase, the field-free results indicate that the OAM projections
$\Delta m_l = 1, 2,$ and $3$ exhibit nearly identical angular distributions with pronounced peaks around $\theta_e = \pm 180^\circ$, while the $\Delta m_l = 2$ yields the smallest TDCS magnitude. In contrast, the $\Delta m_l = 4$ contribution displays the largest TDCS, characterized by a two-peak structure consisting of a forward peak near $\theta_e = 15^\circ$ and a n
dominant backward peak at $\theta_e = \pm 180^\circ$.

In the presence of the laser field, a distinctive behavior emerges when the two Bessel
beams are in phase. For both the $\mathrm{LP}$ and $\mathrm{CP_Y}$ laser-assisted collisions, the odd OAM projections $\Delta m_l = 1$ and $3$ produce identical angular distributions of the TDCS, as evidenced by the complete overlap of the corresponding curves (green solid and red dashed curves) in Figures~\ref{fig:8}(b) and \ref{fig:8}(c). This indicates that, under coherent in-phase superposition, the laser-assisted dynamics become insensitive to the odd value of $\Delta m_l$. However, despite this similarity between $\Delta m_l = 1$ and $3$ within each polarization scheme, the angular profiles of the $\mathrm{LP}$ and $\mathrm{CP_Y}$ configurations differs markedly (see green solid and red-dashed curve in Figures \ref{8}(b) and (c)), reflecting the distinct symmetry and coupling characteristics associated with linear and circular laser polarizations. 
Furthermore, the even OAM projections reveal a significant dependence on the specific laser polarization state, highlighting a clear departure from the behavior of the odd projections. In the LP configuration, $\Delta m_l = 2$ distribution (blue dotted curve) is characterized by a distinct two-peak structure located at $\theta_e = -60^\circ$ and $120^\circ$. In contrast, the $\Delta m_l = 4$ contribution (purple dash-dotted curve) exhibits both forward and backward peaks at $\theta_e = 0^\circ$ and $\theta_e = \pm 180^\circ$, respectively, with the backward peak being dominant over the forward emission.
The $\mathrm{CP_Y}$ case presents a markedly different trend. For $\Delta m_l = 2$, the angular distribution is restricted primarily to a single backward peak, suppressing the forward emission seen in the LP case. Meanwhile, for $\Delta m_l = 4$ TDCS$_{av}$ displays a dual-peak structure exhibiting both forward and backward components. Notably, in direct contrast to the LP results, the forward peak in the $\mathrm{CP_Y}$ configuration dominates the backward emission. This inversion of peak dominance indicates that the circular polarization enhances forward scattering for higher-order OAM transfers, a feature that could be instrumental in controlling ejected electron emission directions in twisted-beam experiments.

As the relative phase between the two Bessel beams is increased from
$\alpha = 0^\circ$ to $60^\circ$ and $90^\circ$, the TDCS exhibits a phase sensitivity that depends on both the OAM projection and the laser polarization.
The phase variation (from $\alpha = 0^\circ$ to $\alpha = 60^\circ$ and $\alpha = 90^\circ$) breaks the similarity of the TDCS observed for odd OAM projections at $\alpha = 0^\circ$ (see green solid and red dashed curves in Figure~9(b) and (c)).
 In the $\mathrm{LP}$ laser-assisted case, the angular distributions for $\Delta m_l = 1$ and $3$, which are identical at $\alpha = 0^\circ$, become distinct for $\alpha = 60^\circ$ and $90^\circ$, while the $\Delta m_l = 1$ and $2$ channels exhibit similar angular profiles differing primarily in magnitude. This indicates that, for $\mathrm{LP}$ scenario, the phase variation modifies the interference between OAM projections, leading to a redistribution of the TDCS without significantly altering the overall emission distribution.
For the $\mathrm{CP_Y}$ configuration, the phase dependence is more intricate. For $\alpha = 60^\circ$, the angular distributions corresponding to the $\Delta m_l$ = 1,2,3 and 4 values remain broadly similar, with the primary differences appearing in the magnitude of the TDCS rather than in the peak positions. When the phase difference is increased to $\alpha = 90^\circ$, a stronger dependence on the $\Delta m_l$ becomes evident, with the $\Delta m_l = 3$ component exhibiting a noticeable modification in both magnitude and angular profile, while the $\Delta m_l = 1, 2,$ and $4$ components remain comparatively closer to their distributions at $\alpha = 60^\circ$. This behavior indicates that increasing the phase difference enhances the interference between the twisted components, leading to a more pronounced redistribution of the TDCS. These observations demonstrate that the relative phase between the superposed twisted beams acts as an additional parameter for controlling the angular distribution of the ionization process.

 		\section{Conclusion}
        We have investigated laser-assisted (e,2e) ionization of hydrogen atom by twisted electron beams, for both linearly and circularly polarized laser fields. Our work focused on the roles of laser polarization, orbital angular momentum (OAM), beam opening angle, and coherent beam superposition on the angualr distribution of TDCS and TDCS$_{av}$. Within the one-photon absorption regime, we find that the presence of a laser field strongly suppresses the triple differential cross section (TDCS), with linear polarization producing a significantly larger reduction than that for the circular polarization.
        A key result is the emergence of an effective kinematic symmetry when the projectile scattering angle ($\theta_s$) equals the opening angle ($\theta_p$) of the twisted electron beam. Under this condition, the angular distribution in the circularly polarized case closely resembles that of the field-free process, while pronounced deviations occur when this geometric relation is not satisfied ($\theta_s \neq \theta_p$). Increasing the OAM of the TEB projectile leads to a systematic decrease in the TDCS magnitude, although the angular distribution remains largely unchanged.
        For laser-assisted collisions, the angular distributions exhibit a clear dependence on the odd and even nature of the OAM at small opening angles ($\theta_p = 1^\circ$ and 4$^\circ$), reflecting the phase structure of the twisted electron beam. This behavior disappears at comparatively large opening angle $\theta_p = 15^\circ$, where the distributions become strongly OAM dependent. For the coherent superposition of two Bessel beams, the in-phase configuration ($\alpha = 0^\circ$) yields identical TDCS distributions for the odd OAM projections. While for a non-zero phase difference ($\alpha = 60^\circ$ and $90^\circ$) breaks this similarity, and the resulting distributions depend on both the OAM projection and the laser polarization. The effect is more substantial for circular polarization, with greater sensitivity to the relative phase between the beams.
   Our results demonstrate that laser polarization, beam geometry, and OAM provide powerful and complementary control parameters for tailoring electron-impact ionization dynamics with twisted electron beams.
		\label{conc}
    	\nocite{apsrev41Control}
			\bibliography{MS_CRSV2}  
		\end{document}